\documentclass[journal]{IEEEtran}

\usepackage[latin1]{inputenc}
\usepackage{xcolor}
\usepackage{changepage}

\usepackage{graphicx}
\usepackage[outdir=./]{epstopdf}
\usepackage{algorithm}
\usepackage{bm}
\usepackage[noend]{algpseudocode} 
\makeatletter 
\def\BState{\State\hskip-\ALG@thistlm}
\makeatother
\newcommand\StateX{\Statex\hspace{\algorithmicindent}}
\usepackage{amsmath,amssymb}
\usepackage{hyperref}
\usepackage{mathtools}
\usepackage{tikz}                    
\usetikzlibrary{shapes,arrows}
\usetikzlibrary{automata,calc,matrix}
\usepackage{xspace}                 
\usepackage{cite}                    

\usepackage{comment}
\usepackage{color}
\usepackage{dashrule}
\usepackage{url}                     
\usepackage{multirow}				
\usepackage{colortbl}

\usepackage{soul}

\newcommand{\ignore}[1]{}

\newtheorem{theorem}{Theorem}[section]
\newtheorem{lemma}[theorem]{Lemma}

\newtheorem{example}[theorem]{Example}

\newtheorem{definition}[theorem]{Definition}

\newtheorem{proposition}[theorem]{Proposition}

\begin{document}
\title{Integer Ring Sieve for Constructing Compact QC-LDPC Codes with Girths 8, 10, and 12}

\author{
\IEEEauthorblockN{Alireza Tasdighi and Emmanuel Boutillon \IEEEmembership{Senior Member,~IEEE} \\}
\IEEEauthorblockA{Universit\'e Bretagne-Sud,\\ Lab-STICC, UMR 6285, CNRS -- Lorient, France\\
\url{ar.tasdighi@gmail.com,} \url{emmanuel.boutillon@univ-ubs.fr}\\}
}

\maketitle

\tikzstyle{every picture}+=[remember picture]
\tikzstyle{na} = [baseline=-.5ex]

\begin{abstract}
This paper proposes a new method of constructing compact fully-connected Quasi-Cyclic Low
Density Parity Check (QC-LDPC) codes with girth $g$ = 8, 10, and 12. The originality of the proposed method is to impose constraints on the exponent matrix $\mathbf{P}$ to reduce the search space drastically. For a
targeted lifting degree of $N$, the first step of the method is to sieve the integer ring $\mathbb{Z}_N$ to make a particular sub-group with specific properties to construct the second column of $\mathbf{P}$ (the first column being filled with zeros). The remaining columns of $\mathbf{P}$ are determined recursively as multiples of the second column by adapting the sequentially multiplied column (SMC) method whereby a controlled greedy search is applied at each step. The codes constructed with the proposed semi-algebraic method show lengths that can be significantly shorter than their best counterparts in the literature.
\end{abstract}
\begin{IEEEkeywords}
QC-LDPC Code Construction, Girth, Multiplicative Group, Cyclic Subgroup, Greedy Search Method.
\end{IEEEkeywords}
\section{Introduction} 
\label{sec:intro}
It has been more than two decades since the rediscovery of low-density parity-check (LDPC) codes as a class of modern channel coding \cite{MacKay1999}. Quasi-cyclic (QC) LDPC codes, a special class of LDPC codes, enable efficient parallel hardware implementation and have been adopted in many communication standards. Examples include the WIFI standard \cite{WirelessStandard2017}, digital video broadcasting (DVB) standard \cite{DVB2009}, CCSDS standards \cite{CCSDS2015}, and more recently, the 5G standard \cite{3GPP2019}. The Performance of an LDPC iterative decoder is significantly impacted by the minimum size of a cycle of the Tanner graph (the so-called girth). Maximizing the girth for a given code size and code rate, or reciprocally, finding the minimum code size for a given code rate and girth size, has proved challenging for the past two decades \cite{MFossorier1,sullivan1,Wang2008,RAM2011,DSMB2012,Bocharova1,JLR2013,MDSB2015,Gholami2017,ATasdighi1,ATasdighi2,XYYZ2018,MBATMB2018,AEA2018,SA2018}.
This paper is focused on the construction of a QC-LDPC matrix constructed from a fully connected exponent matrix. QC-LDPC codes can be used on their own for applications requiring a very low Frame Error Rate (e.g. memory storage). They can also be combined with a masking technique to build good irregular QC-LDPC codes \cite{Masking07, Masking13, Masking14}. 
In addition, it has recently been shown that by using some spreading techniques, a class of SC-QC-LDPC convolutional codes with very low syndrome memory could be constructed based on QC-LDPC codes constructed from a fully-connected exponent matrix \cite{MBATMB2018,Felt1,MBAT2018,MHTAT2018,MBMBFC2021}.
Specifically, \cite{MBAT2018} asserts that given a fixed girth and
degree distribution, the smaller the lifting degree of QC-LDPC codes, the smaller the size of the syndrome memory of SC-QC-LDPCC codes and this results in better performance of such code under windowed decoding. 
This study focuses solely on constructing short length QC-LDPC codes with girth $g=8$, $10$, and $12$. However,
we keep in mind that SC-QC-LDPCC codes are potential candidates for beyond 5G applications, and good QC-LDPC codes are the basis of good SC-QC-LDPC codes.

QC-LDPC codes can be divided into two major classes: 1) random-like codes constructed by means of a computer search under efficient algorithms, and 2) structured codes constructed using algebraic tools \cite{JSKWD2017}. Individually, these methods all have drawbacks. Search-based methods (even heuristic or exhaustive ones) require high search complexity but may find codes with shorter length than the ones obtained with algebraic methods. In contrast, algebraic methods determine the code explicitly (e.g. array code \cite{JSKWD2017} of girth $6$) but to date,  algebraic methods are known to construct small-girth codes only, not high-girth codes. Defining algebraic properties that are perfectly matched with high-girth conditions resulting in the explicit construction of short-length codes is one of the main shortcomings of algebraic methods. In this paper, we try to combine the two methods in order to construct large-girth QC-LDPC codes with a short length in considerably lower search complexity. We take the search-based sequentially multiplied column (SMC) construction method \cite{MHTAT2018} as our search algorithm and modify it by introducing an algebraic property for the second column of the  exponent matrix of the code. The second column with the asserted algebraic property is found by an Integer Ring Sieve (IRS) method in a way that eventually leads to reducing the search space. As a result, a  semi-algebraic fast search-based method of constructing high-girth QC-LDPC codes is proposed, and many constructed codes of girth $g=8$, $10$, and $12$ with different rates and degrees are reported. Most of the constructed codes have lengths shorter than (by up to 35\%), or equal to, their counterparts in the literature.  
 The paper also proposes matrices with sizes not yet reported in the literature. 

The rest of the paper is organized as follows: Section II presents earlier results regarding SMC construction-based QC-LDPC codes. Section III presents the proposed IRS construction method. Numerical results are provided in Section IV. 

\section{Preliminaries} \label{sec:preleminaries}

In this section, we review the construction of Quasi-Cyclic LDPC codes. Next we discuss the conditions that result in QC-LDPC codes with good topological properties. 
\subsection{QC-LDPC block codes}
\label{subsec:QC-LDPC block codes}
Let us consider a \textit{fully-connected} QC-LDPC block code in which the parity-check matrix is an $m \times n$ array of $N \times N$ circulant permutation matrices (CPMs), $\mathbf{I}(p_{ij})$, $0 \leq i \leq m - 1$, $0 \leq j \leq n - 1$, where $N$ is the \textit{lifting degree} of the code. $\mathbf{I}(p_{ij})$ is obtained from the identity matrix through a cyclic shift of its rows by $p_{ij}$ positions, with $0 \leq p_{ij} \leq N - 1$. The code length is $L=nN$, the column degree of the parity-check matrix is presented by $m$ and the row degree of the parity-check matrix is presented by $n$\footnote{In the case that the QC-LDPC code is not fully-connected, $m$ and $n$ are often noted by $d_{v}$ and $d_{c}$ in the literature, respectively.}. The resulting rate of the code is $r \geq 1 - m/n$. The $m \times n$ matrix $\mathbf{P}$ having the integer values $p_{ij}$ as its entries is referred to as the \textit{exponent matrix} of the code.

\begin{definition}\label{def:cycle} A cycle $C_{2k}$ of length $2k$ in the exponent matrix $\mathbf{P}$ is defined as an ordered set of $2k$ positions  $\{(m_{s}, n_{s})\}_{s=0, 1, ..., 2k-1}$ in the matrix $\mathbf{P}$ satisfying the following three conditions: 1) $m_{s} = m_{s+1}; n_{s} \neq n_{s+1}$ when $s$ is even, 2) $m_{s} \neq m_{s+1}; n_{s} = n_{s+1}$ when $s$ is odd, 3) additions in the indexes are done modulo $2k$, i.e. when $s = 2k-1$, $n_{s+1} = n_{2k} = n_0$. For example, the length-6 cycle $\mathcal{C}_6 = \{(0,0), (0,1), (1,1), (1,2), (2,2), (2,0)\}$ is represented in Fig. \ref{fig:CyclePaths}.c.
\end{definition}

\begin{definition} For a given exponent matrix $\mathbf{P}$, we define $\theta_{\mathbf{P}}$ as a function that associates to a cycle $\mathcal{C}_{2k}$ the integer $\theta_{\mathbf{P}}(\mathcal{C}_{2k})$ as

\begin{equation}
 \theta_{\mathbf{P}}(\mathcal{C}_{2k}) = \sum_{s=0}^{2k-1} (-1)^s p_{m_{s}n_{s}}. 
\label{eq:Fossorier1}
\end{equation}  

\end{definition}
According to \cite{MFossorier1}, a necessary and sufficient condition for the existence of a cycle of length $2k$ in the Tanner graph of the QC-LDPC block code is the existence of a cycle $\mathcal{C}_{2k}$ in $\mathbf{P}$ satisfying $\theta_{\mathbf{P}}(\mathcal{C}_{2k}) = 0 \mod N$.
Thus, a necessary and sufficient condition to ensure a QC-LDPC code of girth $g$ is that all cycles $\mathcal{C}_{2k}$, $2k<g$ of the exponent matrix $\mathbf{P}$ satisfy

\begin{equation}
\theta_{\mathbf{P}}(\mathcal{C}_{2k}) \neq 0 \mod N. 
\label{eq:Fossorier}
\end{equation}   

We define an {\em inevitable cycle} of length $2k$ to be a cycle $\mathcal{C}^i_{2k}$ so that $\theta_{\mathbf{P}}(\mathcal{C}^i_{2k}) = 0$  regardless of what the values of $p_{ij}$s are. In \cite{MFossorier1} it is shown that fully-connected CPM-based QC-LDPC codes always contain inevitable cycles of length $12$, and thus their girth cannot be larger than $12$.

\subsection{Code design using sequentially multiplied columns}
\label{subsec:sequentially multiplied columns}

Searching for an exponent matrix $\mathbf{P}_{m \times n}$ of a given girth $g=2k$ and a lifting degree $N$ is a complex task. First, the raw number of exponent matrices is exponential in $m$ and $n$ ($N^{mn}$ exactly); second, the number of cycles $C_{2k}$ of length $2k = g-2$ (and thus, the number of equations (\ref{eq:Fossorier}) to satisfy), increases in the order of $\mathcal{O}(m^{k} n^{k})$. In fact, a cycle of length $2k$ can involve up to $k$ columns and $k$ rows.
Solutions with reduced complexity were proposed in \cite{Gholami2017} and \cite{ATasdighi2}, but the corresponding design methods result in girth $g = 8$. For constructing short codes with higher girths (i.e. $g=10$, $12$), many methods have been developed. To the best of the authors' knowledge, the results in \cite{MHTAT2018} for QC-LDPC codes with girth $g=10,12$ found by applying the SMC construction technique are the shortest ones in the literature. Let us recall the basic assumptions of the design method proposed in \cite{MHTAT2018}. The design of the parity-check matrix of a QC-LDPC block code with lifting degree $N$ starts from an exponent matrix with the following form (SMC assumption):

\begin{equation}
\mathbf{P}_{m\times n}=\left[\begin{array}{c|c|c|c|c|c}
\vec{0} & \vec{P}_{1} & \gamma_{2} \otimes \vec{P}_{1} & \gamma_{3} \otimes \vec{P}_{1} & \ldots &  \gamma_{n-1} \otimes \vec{P}_{1}
\end{array}\right],
\label{eq:SMCexpomatrix}
\end{equation}
with $m$, $n$, $\in \mathbb{N}$, $m < n$, and $\vec{0}$ and $\vec{P_1}$ being
$m \times 1$ column vectors with entries in $\lbrace 0,1,\cdots, N-1\rbrace$. The
vector $\vec{0}$ is filled with all zero entries, while the entries of
the vector $\vec{P_1}$ are chosen as follows: the first entry is zero, the second entry is one, and the other entries are chosen in
$\lbrace 2, 3, \cdots ,N-1\rbrace$. Then, the subsequent
vectors have the form $\gamma_j \otimes \vec{P_1}$ ($j = 2,3,\cdots,n-1$), where
$\otimes$ denotes multiplication modulo $N$ of each term of $\vec{P_1}$ with $\gamma_j \in \lbrace 2, 3,\cdots,N-1\rbrace$, $\gamma_j < \gamma_{j+1}$.

To achieve this result, the authors of \cite{MHTAT2018} established a recursive greedy search algorithm (see algorithm 1 in \cite{MHTAT2018}) to determine, for a large enough lifting factor $N$, the exponent matrix  $\mathbf{P}^{SMC}_{m \times n}$  that satisfies (\ref{eq:Fossorier}) for all cycles of girth lower than $g$. 
This search algorithm is supposed to find $n-2$ (resp., $m-2$) non-zero and distinct elements to be placed in the second row (resp., column) of $\mathbf{P}^{SMC}_{m\times n}$. These elements vary from $1$ to $N-1$, and, in the worst case, 
the overall possibilities are equal to $\binom{N-1}{n-2}\binom{N-1}{m-2}$. 
For high-rate and high-girth codes,  the lifting degree is much bigger than $m$ and $n$ (i.e. $m,n\ll N$), and the whole search space is of order $\mathcal{O}\left(\left(N-1\right)^{m+n-4}\right)$. It has to be noted that for a desired girth $g$, each realization of the matrix $\mathbf{P}^{SMC}_{m\times n}$ requires checking all the constraints of type (\ref{eq:Fossorier}) with $k<g/2$. 

In the next section, an improved construction method is proposed.  

\section{Integer Ring Sieve to find permissible elements for the vector $\vec{P}_1$}
\label{sec:Integer Ring Sieve}

This section is divided into four parts. In Part A, we propose the definition of  \textit{strictly equivalent} relations between cycles of a fully connected exponent matrix $\mathbf{P}$ based on (\ref{eq:Fossorier}). Next, we give a theorem for counting the number of strictly equivalent classes of length $2k$ ($k=2,3,4,5$) in $\mathbf{P}$. In Part B, we show that a careful selection of the second column of matrix $\mathbf{P}_{m\times n}$ (i.e. $\vec{P}_1$) can create a new type of equivalence relation called ``Integer-Ring (IR) equivalence''. When combining strict equivalence and IR-equivalence together, the number of equivalent classes is divided by a factor close to 3 for $m = 3$ and close to $m - 1$ for $m \geq 4$. In Part C, we show results regarding the existence of integer rings with a property defined in Part B. Our greedy search algorithm is explained in Part D with a pseudocode. Complexity analyses to highlight the important role of our sieving method in reducing the search space are also provided in this final part. 
\subsection{Equivalent relations between the set of cycles}
\begin{definition}[strictly equivalent cycle]\label{definition:equivalent cycle}
Two cycles $\mathcal{C}$ and $\mathcal{C}'$ of an exponent matrix $\mathbf{P}$ are said to be strictly equivalent if and only if $\theta(\mathcal{C}) = 0 \Leftrightarrow \theta_\mathbf{P}(\mathcal{C}') = 0$ regardless of the value of $\mathbf{P}$. This relation of equivalence defines equivalent classes. By convention, if two cycles of different lengths are equivalent, only the cycle with the smallest length will be considered. 
\end{definition}

\begin{figure*}
\centering
\includegraphics[scale=0.46]{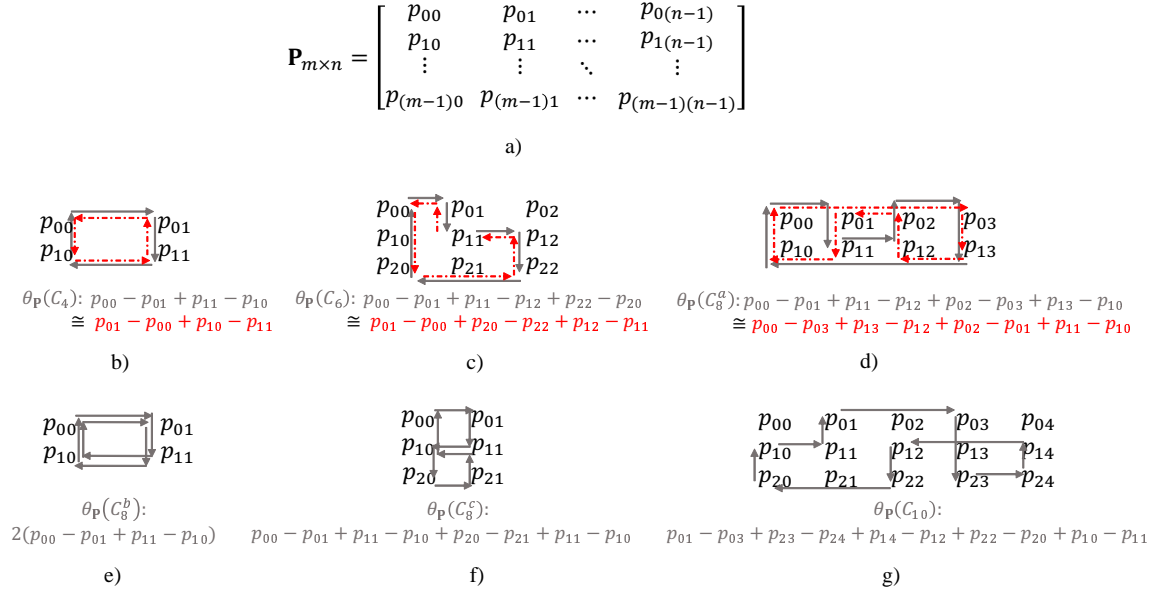}
\caption{Sample paths for cycles of length from $4$ to $10$ involved in an exponent matrix $\mathbf{P}$: a) exponent matrix of size $m\times n$. b) paths of length $4$ of strictly equivalent cycles. c) paths of length $6$ of strictly equivalent cycles. d) paths of length $8$ of strictly equivalent cycles. e) path of a length $8$ cycle non-strictly equivalent to (d). f) another path of a length $8$ cycle non-strictly equivalent to the paths (d) and (e). g) path of a length $10$ cycle.}
\label{fig:CyclePaths}
\end{figure*}

Fig. \ref{fig:CyclePaths} gives several examples of strictly equivalent cycles in an exponent matrix $\mathbf{P}$.

\begin{definition}[number of equivalent classes]
\label{definition:equivalent class}
The number of classes of cycles of length $2k$ on an $m \times n$ exponent matrix is denoted as  $\#\mathcal{C}^{m,n}_{2k}$. Cycles in each class are strictly equivalent.  
\end{definition}

\begin{definition}[Cycle's tracking matrix of order $2k$]
\label{definition:cycle's tracking matrix}
\textit{The cycle's tracking matrix of order $2k$} is a square matrix $T^{\mathcal{C}_{2k}}$ of size $k$ ($k=2,3,\cdots$) where its ($i$, $j$)$^{\text{\small th}}$ component counts the number of classes (for the strictly equivalent relation) of length $2k$ that involve all rows and columns of a matrix of size $i\times j$. 
\end{definition}

The matrices $T^{\mathcal{C}_{4}}$, $T^{\mathcal{C}_{6}}$, $T^{\mathcal{C}_{8}}$ and $T^{\mathcal{C}_{10}}$ are respectively 

\begin{align*}
\begin{small}
T^{\mathcal{C}_4}=\begin{bmatrix}
0&0\\
0&1
\end{bmatrix},\;
T^{\mathcal{C}_6}=\begin{bmatrix}
0&0&0\\
0&0&0\\
0&0&6
\end{bmatrix}
\end{small}
\end{align*}

\begin{align*}
\begin{small}
T^{\mathcal{C}_8}=\begin{bmatrix}
 0 & 0 & 0  & 0 \\
 0 & 1 & 3  & 3 \\
 0 & 3 & 18 &36 \\
 0 & 3 & 36 &72 
\end{bmatrix},\;
T^{\mathcal{C}_{10}}=\begin{bmatrix}
 0 & 0 & 0  & 0  & 0\\
 0 & 0 & 0  & 0  & 0\\
 0 & 0 & 60 &180 & 180\\
 0 & 0 &180 &900 & 1440\\
 0 & 0 &180 &1440 &1440
\end{bmatrix}.
\end{small}
\end{align*}

It has to be noted that $T^{\mathcal{C}_{2k}}$ is symmetrical (i.e. $T^{\mathcal{C}_{2k}}=\left(T^{\mathcal{C}_{2k}}\right)^{T}$) as the number of non-equivalent cycles involved in a $i\times j$ matrix is equal to the number of such cycles involved in a matrix of size $j\times i$. Note that the matrices $T^{C_{4}}$ to $T^{C_{10}}$ have been obtained using a computer program. 

In an exponent matrix $\mathbf{P}_{m \times n}$ of size $(m\times n)$, the total number $\#\mathcal{C}_{2k}^{m,n}$ of non-equivalent cycles of girth $2k$ can be computed as the number of distinct sub-matrices $(i,j)$ compatible with a cycle of length $2k$ (i.e. $2 \leq i \leq k$ and $2 \leq j \leq k$) multiplied by $T^{\mathcal{C}_{2k}}(i,j)$, this results in
\begin{equation}
\label{theorem:nonequivalent potential cycles general case}
\#\mathcal{C}^{m,n}_{2k}=\sum\limits_{i=2}^{\min \lbrace k,m\rbrace}\sum\limits_{j=2}^{\min \lbrace k,n\rbrace} T^{\mathcal{C}_{2k}}(i,j) {m\choose i} {n\choose j}, 
\end{equation}
\noindent where ${n \choose r}$ represent the binomial coefficient $\frac{n!}{r!(n-r)!}$. Table \ref{table:nonequivalent potential cycles} gives the values of $\#\mathcal{C}^{m,n}_{2k}$ for $k=2,3,4,5$, $2\leq m\leq 5$ and $2\leq n\leq 10$. The total number of strictly equivalent cycles of length lower than $g$ of an $m \times n$ exponent matrix $\mathbf{P}_{m\times n}$ is thus given as 
\begin{equation}
\label{theorem:total number}
\#\mathbf{P}^{m \times n}(g) =\sum\limits_{k=2}^{g/2-1} \#\mathcal{C}^{m,n}_{2k}. 
\end{equation}
For example, if $g=12$, $m=3$ and $n=10$, then satisfying that a given exponent matrix $\mathbf{P}_{3,10}$ of size $(3,10)$ generates a girth 12 QC-LDPC code with a given lifting degree $N$ implies checking a total of $\#\mathbf{P}^{3,10}(12) = 135+720+12960+90360= 104175$ equations. To reduce the number of non-equivalent classes, some constraint can be added to the exponent matrix. 

\begin{table*}[htb]
\renewcommand{\arraystretch}{1}
\caption{Number of strictly equivalent classes of length $2k$ cycles ($k=2,3,4,5$) that are involved in matrix $\mathbf{P}_{m \times n}$, when, $2\leq m\leq 5$ and $2\leq n\leq 10$.}
\label{table:nonequivalent potential cycles}
\centering
\begin{tabular}{|@{}c@{}||@{}c@{}|@{}c@{}|@{}c@{}|@{}c@{}||@{}c@{}|@{}c@{}|@{}c@{}|@{}c@{}||@{}c@{}|@{}c@{}|@{}c@{}|@{}c@{}||@{}c@{}|@{}c@{}|@{}c@{}|@{}c@{}|}
\hline
$-$ & \multicolumn{4}{c||}{$m=2$} & \multicolumn{4}{c||}{$m=3$} & \multicolumn{4}{c||}{$m=4$} & \multicolumn{4}{c|}{$m=5$} \\ \hline
$-$ & $\#\mathcal{C}^{2,n}_{4}$ & $\#\mathcal{C}^{2,n}_{6}$ & $\#\mathcal{C}^{2,n}_{8}$ & $\#\mathcal{C}^{2,n}_{10}$ & $\#\mathcal{C}^{3,n}_{4}$ & $\#\mathcal{C}^{3,n}_{6}$ & $\#\mathcal{C}^{3,n}_{8}$ & $\#\mathcal{C}^{3,n}_{10}$  & $\#\mathcal{C}^{4,n}_{4}$ & $\#\mathcal{C}^{4,n}_{6}$ & $\#\mathcal{C}^{4,n}_{8}$ & $\#\mathcal{C}^{4,n}_{10}$  & $\#\mathcal{C}^{5,n}_{4}$ & $\#\mathcal{C}^{5,n}_{6}$ & $\#\mathcal{C}^{5,n}_{8}$ & $\#\mathcal{C}^{5,n}_{10}$  \\ \hline \hline
$n=2$ & $1$ & $0$ & $1$ & $0$ & $3$ & $0$ & $6$ & $0$ & $6$ & $0$ & $21$ & $0$ & $10$ & $0$ & $55$ & $0$ \\ \hline
$n=3$ & $3$ & $0$ & $6$ & $0$ & $9$ & $6$ & $45$ & $60$ & $18$ & $24$ & $189$ & $420$ & $30$ & $60$ & $555$ & $1680$ \\ \hline
$n=4$ & $6$ & $0$ & $21$ & $0$ & $18$ & $24$ & $189$ & $420$ & $36$ & $96$ & $864$ & $3300$ & $60$ & $240$ & $2640$ & $14460$ \\ \hline
$n=5$ & $10$ & $0$ & $55$ & $0$ & $30$ & $60$ & $555$ & $1680$ & $60$ & $240$ & $2640$ & $14460$ & $100$ & $600$ & $8200$ & $65940$ \\ \hline
$n=6$ & $15$ & $0$ & $120$ & $0$ & $45$ & $120$ & $1305$ & $4980$ & $90$ & $480$ & $6345$ & $45660$ & $150$ & $1200$ & $19875$ & $212340$ \\ \hline
$n=7$ & $21$ & $0$ & $231$ & $0$ & $63$ & $210$ & $2646$ & $12180$ & $126$ & $840$ & $13041$ & $116760$ & $210$ & $2100$ & $41055$ & $548940$ \\ \hline
$n=8$ & $28$ & $0$ & $406$ & $0$ & $84$ & $336$ & $4830$ & $26040$ & $168$ & $1344$ & $24024$ & $257880$ & $280$ & $3360$ & $75880$ & $1220520$ \\ \hline
$n=9$ & $36$ & $0$ & $666$ & $0$ & $108$ & $504$ & $8154$ & $50400$ & $216$ & $2016$ & $40824$ & $511560$ & $360$ & $5040$ & $129240$ & $2431800$ \\ \hline
$n=10$ & $45$ & $0$ & $1035$ & $0$ & $135$ & $720$ & $12960$ & $90360$ & $270$ & $2880$ & $65205$ & $934920$ & $450$ & $7200$ & $206775$ & $4457880$ \\ \hline
\end{tabular}
\end{table*}

\subsection{Integer Ring-based construction of $\vec{P}_{1}$} \label{subsec:designating second column}
Let us recall the usual notations used in integer ring theory \cite{Hans1994}. The greatest common divisor of $a$ and $b$ is noted $\gcd\left(a, b\right)$. The ring of integers modulo $N$ is denoted by $\mathbb{Z}_N$. The multiplicative group of integers modulo $N$ (set of values of $\mathbb{Z}_{N}$ coprime with $N$) is denoted by $\mathbb{Z}_{N}^{\times}$. An element $a \in \mathbb{Z}_{N}^{\times}$ generates a finite cyclic subgroup $\langle a\rangle=\lbrace 1,a, a^2, a^3,  \cdots, a^{O_N(a)-1}\rbrace$, with $O_N(a)$ the smallest non-null integer satisfying $a^{O_N(a)} = 1 \mod N$. The value $O_N(a)$ is called the order of $a$ in $\mathbb{Z}_{N}^{\times}$.

In this part, we try to pick the non-zero elements of $\vec{P}_{1}$ (see (\ref{eq:SMCexpomatrix})) from a specific cyclic subgroup of $\mathbb{Z}^{\times}_{N}$. Depending on the value of $m$, we propose to allocate some or all of the elements in this subgroup to $p_{j1}$ ($1\leq j\leq m-1$). The main reason behind such allocation is to add to the strictly equivalent cycles a new type of equivalent cycles that are called ``Integer Ring  equivalent'' (IR-equivalent) cycles. Reducing the number of equivalent cycles (i.e. the number of equations that should be satisfied) has several positive side effects. First, it directly accelerates the research algorithm since the number of constraints to be checked is reduced. Second, it reduces the search space: because of the a priori selection of the  $\vec{P}_{1}$ column coefficients, only the determination of the elements $\gamma_{j}$ ($j=2,3,\cdots,n-1$) is required. Third, and above all, it increases the likelihood of finding a solution. The question is open whether, for a given girth and exponent matrix size, the smallest lifting factor can be obtained only with IRS exponent matrices. Note that, by serendipity, the authors forgot to kill an unconstrained search based on the method proposed in \cite{AEA2018} to generate an $(m,n) = (3,7)$  exponent matrix with a lifting degree $N=133$. After five months of silent computation, a solution was found, which involved an exponent matrix with the IRS structure (see Appendix I).

\begin{definition} A type-I IRS QC-LDPC matrix is a matrix obtained from the full SMC exponent matrix $\mathbf{P}^{a,I}$ of size $m \times n$ with vector column $\vec{P}_1$ defined as $\vec{P}_1 = (0, 1, a, a^2, \ldots , a^{m-1})^T$, with $a$ an element of $\mathbb{Z}^\times_N$ of order $O_N(a) = m$ and $N$ representing the lifting degree.
\end{definition}

\begin{definition} A type-II IRS QC-LDPC matrix is a matrix obtained from the full SMC exponent matrix $\mathbf{P}^{a,II}$ of size $(3,n)$ with vector column $\vec{P}_1$ defined as $\vec{P}_1 = (0, 1, a)^T$, with $a$ an element of $\mathbb{Z}^\times_N$ satisfying $a(1-a) = 1 \mod N$ and $N$ representing the lifting degree.
\end{definition}

The element $(i,j)$ of the type-I IRS exponent matrix $\mathbf{P}^{a,I}$ is defined as 
\begin{equation}
\mathbf{P}^{a,I}(i,j)= u(i)a^{i-1}\gamma_j.
\label{eq:P^a,I}
\end{equation}
 \noindent with $u(i) = 0$ if $i=0$, 1 otherwise. Note that the elements of a type-II IRS exponent matrix $\mathbf{P}^{a,II}$ have also the same form as (\ref{eq:P^a,I}).   

Let $\pi$ be a permutation on the number of rows $(0,1, \ldots, m-1)$ of an exponent matrix $\mathbf{P}_{m\times n}$. The permutation $\pi$ is defined in the sequel by the vector $(\pi(0), \pi(1), \ldots, \pi(m-1))$.  
From the permutation $\pi$, $\pi(\mathcal{C}_{2k})$ is defined as the function that is associated to a cycle $\mathcal{C}_{2k} = \{(m_s, n_s)\}_{s = 0, 1, \ldots, 2k-1}$ of $\mathbf{P}_{m\times n}$ the cycle of same length $\pi(\mathcal{C}_{2k})$ of $\mathbf{P}_{m\times n}$ defined as $\pi(\mathcal{C}_{2k}) = \{(\pi(m_s), n_s)\}_{s = 0, 1, \ldots, 2k-1}$.

Note that, in the general case, cycles $\mathcal{C}_{2k}$ and $\pi(\mathcal{C}_{2k})$ are not strictly equivalent. 
\begin{definition} Let $\pi^{I,1}$ be the permutation over the ordered vector $(0, 1, \ldots, m-1)$ defined as $\pi^{I,1} = (0, 2, 3, \ldots, m-1, 1)$ and $\pi^{I,l}$ the rotation obtained by applying the permutation $\pi^{I,1}$ $l$ times. Thus, for $l = 1, 2, \ldots, m-1$, we have $\pi^{I,l}(0) = 0$ and $\pi^{I,l}(i) = ((i+l) \mod m) + ((i+l) \div m)$. Note that $(i+l) \div m$, the Euclidean division of $(i+l)$ by $m$ takes the value 0 when $(i+l) < m$, and 1 when $m \leq i+l < 2m$.  
\end{definition}
\begin{theorem} \label{theorem:cyclic subgroup order m}
Let $\mathbf{P}^{a,I}_{m\times n}$ be a type-I IRS exponent matrix. Then, for all cycles $\mathcal{C}_{2k}$ of length $2k$ of $\mathbf{P}^{a,I}_{m\times n}$, $\theta_{\mathbf{P}^{a,I}}(\mathcal{C}_{2k}) = 0$ if and only if $\theta_{\mathbf{P}^{a,I}}(\pi^{I,l}(\mathcal{C}_{2k})) = 0$, $l=1, 2, \ldots m-1$.
\end{theorem}

\begin{IEEEproof} Let $\mathcal{C}_{2k}$ be a cycle of length $2k$ of $\mathbf{P}^{a,I}_{m\times n}$. According to (\ref{eq:Fossorier1}) and (\ref{eq:P^a,I}), 
\begin{equation}
\theta_{\mathbf{P}^{a,I}}(\mathcal{C}_{2k})  = \sum_{s=0}^{2k-1} (-1)^{s} u(m_s) a^{m_s-1}\gamma_{n_s}.
\end{equation}
Since $O_N(a) = m$ in $\mathbb{Z}_N^\times$, then $a^m = 1$ and $a$ and $N$ are coprime. Since $a$ and $N$ are co-prime, $\theta_{\mathbf{P}^{a,I}}(\mathcal{C}_{2k})=0 \mod N$ if and only if $a^l\theta_\mathbf{P}^{a,I}(\mathcal{C}_{2k})=0 \mod N$ for $l = 1, 2, \ldots, m-1$. The equation $a^l\theta_{\mathbf{P}^{a,I}}(\mathcal{C}_{2k})=0 \mod N$ gives 
\begin{equation}
\sum_{s=0}^{2k-1} (-1)^{s} u(m_s) a^{m_s - 1 + l}\gamma_{n_s} = 0 \mod N.
\label{eq:pi}
\end{equation}
Since $a^m = 1$, $u(m_s)a^{m_s + l - 1} = u(\pi^{I,l}(m_s))a^{\pi^{I,l}(m_s)-1}$, thus (\ref{eq:pi}) is equivalent to $\theta_{P^{a,I}}(\mathcal{C}_{2k}) = 0$.
\end{IEEEproof}

\begin{figure*}[htb]
\centering
\includegraphics[scale=0.49]{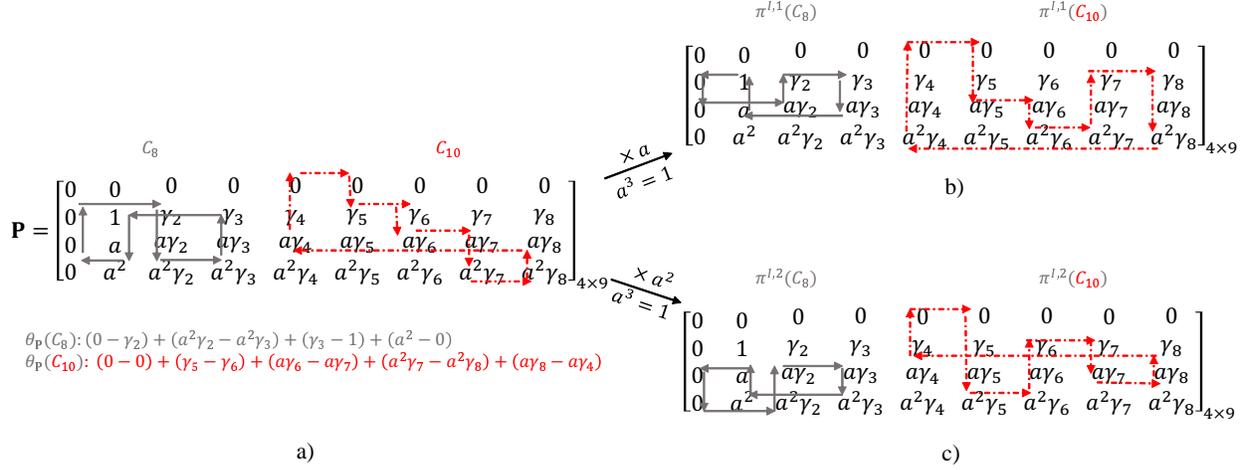}
\caption{Samples of cycles with different length in $\mathbf{P}_{4\times 9}$: a) primary underlined paths for cycles  $\mathcal{C}_8$ and $\mathcal{C}_{10}$. b) isomorphic paths to the paths in part (a) derived from transformation $\pi^{I,1}(\mathcal{C})$. c)  isomorphic paths to the paths in part (a) derived from transformation $\pi^{I,2}(\mathcal{C})$}
\label{fig:CycleReducdv4}
\end{figure*} 

\begin{example}
Suppose that $\mathbf{P}_{4\times 9}$ is the exponent matrix in Fig. \ref{fig:CycleReducdv4} (a), and $\mathcal{C}_{2k}$ ($k=4,5$) are the cycles with the paths depicted in Fig. \ref{fig:CycleReducdv4}.a. According to theorem \ref{theorem:cyclic subgroup order m}, there are two other IR-equivalent cycles to $\mathcal{C}_{2k}$, $\pi^{I,1}(\mathcal{C}_{2k})$ given in Fig. \ref{fig:CycleReducdv4}.b and $\pi^{I,2}(\mathcal{C}_{2k})$ given in Fig. \ref{fig:CycleReducdv4}.c. 
\end{example}

Thus, when a type-I IRS exponent matrix is used, the IR-equivalence between cycles reduces almost by a factor of $m-1$ the number of equivalence classes. The reduction is not exactly $m-1$ because in some rare specific cases, $\pi^{I,l}(\mathcal{C}_{8})$ can also be strictly equivalent to $\mathcal{C}_{8}$. For example, for $(m,n)=(5,2)$, the length-8 cycle $\mathcal{C}_8 = \{(0,0)$, $(0,1)$, $(2,1)$, $(2,0)$, $(0,0)$, $(0,1)$, $(4,1)$, $(4,0)\}$ gives with the permutation $\pi^{I,2} = (0, 3, 4, 1, 2)$ the cycle $\pi^{I,2}(\mathcal{C}_8) = \{(0,0)$, $(0,1)$, $(4,1)$, $(4,0)$, $(0,0)$, $(0,1)$, $(2,1)$, $(2,0)\}$, which is strictly equivalent to $\mathcal{C}_8$ (just a 4 position shift).

\begin{lemma}\label{lemma:O(6)} If $a \in \mathbb{Z}_N^\times$, satisfies $a(1-a) = 1 \mod N$, then $a^3 = -1 \mod N$. Moreover, for $N > 3$, $O_N(a)=6$.
\end{lemma}

\begin{IEEEproof} $a(1-a) = 1 \mod N$ implies $a^2 -a +1  = 0 \mod N$, thus $(1+a)(a^2 -a +1) = 0 \mod N$, and finally, $ (a^3 + 1) =0 \mod N$. Since $a^3 = -1 \mod N$, we can deduce that $O_N(a) \neq 3$ and $a^6 = 1$. The case $a^2 = 1 \mod N$ gives $a = 2$ and $N = 3$. Thus, if $N>3$, we have $O_N(a) = 6$.
\end{IEEEproof}

\begin{definition} Let us define $\pi^{II,1}$ and $\pi^{II,2}$ the permutations $\pi^{II,1}=(1,2,0)$ and $\pi^{II,2} = \pi^{II,1} \circ \pi^{II,1} = (2,0,1)$.
\end{definition}

\begin{theorem} \label{theorem:cyclic subgroup order 3}
Let $\mathbf{P}^{a,II}$ be a type-II IRS exponent matrix. Then, for all cycle $\mathcal{C}_{2k}$ of length $2k$ of $\mathbf{P}^{a,II}$,  for $l=1, 2$, $\theta_{P^{a,II}}(\mathcal{C}_{2k}) = 0$ if and only if $\theta_{P^{a,II}}(\pi^{II,l}(\mathcal{C}_{2k})) = 0$.
\end{theorem}

\begin{IEEEproof} Let $\mathcal{C}_{2k}$ be a cycle of length $2k$ of $\mathbf{P}^{a,II}$. Since $O_N(a) = 6 \mod N$, then according to the definition of a type-II exponent matrix, $a(1-a) = 1 \mod N$ and $a^2 = a-1$. Thus,   $\theta_{P^{a,II}}(\mathcal{C}_{2k})=0 \mod N$ if and only if $a\theta_{\mathbf{P}^{a,II}}(\mathcal{C}_{2k})=0 \mod N$ and if and only if $a^2\theta_{\mathbf{P}^{a,II}}(\mathcal{C}_{2k})=0 \mod N$. 
According to the definition of a cycle, $n_s = n_{s+1}$ when $s$ odd. This property is equivalent to $n_s = n_{s-1}$ when $s$ even, thus it is possible to write $\theta_{\mathbf{P}^{a,II}}(\mathcal{C}_{2k}) =0 \mod N$ as
\begin{equation}
  \sum_{s=0}^{k-1} p_{m_{2s}n_{2s}} - p_{m_{2s-1}n_{2s}} = 0 \mod N.
\label{eq:piII}
\end{equation}
\noindent Thus, using  (\ref{eq:P^a,I}), $\theta_{\mathbf{P}^{a,II}}(\mathcal{C}_{2k}) =0 \mod N$ can be expressed as
\begin{align}
    \begin{split}
\sum_{s=0}^{k-1} (u(m_{2s})a^{m_{2s}-1} - u(m_{2s-1})a^{m_{2s-1}-1}  ))\gamma_{n_{2s}} = \\
0 \mod N.
\end{split}
\label{eq:piII_1}
\end{align}
By taking $\lambda(m_{2s}, m_{2s-1}) = u(m_{2s})a^{m_{2s}-1} - u(m_{2s-1})a^{m_{2s-1}-1}$, (\ref{eq:piII_1}) gives 
\begin{equation}
    \sum_{s=0}^{k-1} \lambda(m_{2s}, m_{2s-1})\gamma_{n_{2s}} = 0 \mod N.
\label{eq:piII2}
\end{equation}
Since $m_{2s} \neq m_{2s-1}$ and $m = 3$, the only six possible couples $(m_{2s}, m_{2s-1})$ are $(0,1)$, $(0, 2)$, $(1,2)$, $(1,0)$, $(2,0)$ and $(2,1)$. Table \ref{typeII} shows that  $a\lambda(m_{2s},m_{2s -1})=-\lambda(\pi^{II,1}(m_{2s}),\pi^{II,1}(m_{2s-1}))$ for all possible couples $(m_{2s}, m_{2s-1})$. Finally, since $\pi^{II,2} = \pi^{II,1} \circ \pi^{II,1}$, $\theta_{\mathbf{P}^{a,II}}(\pi^{II,2}(\mathcal{C}_{2k})) = (-a)^2\theta_{\mathbf{P}^{a,II}}(\mathcal{C}_{2k})$ and thus: 
\begin{align*}
\theta_{\mathbf{P}^{a,II}}(\pi^{II,2}(\mathcal{C}_{2k})) = 0 \Longleftrightarrow\theta_{\mathbf{P}^{a,II}}(\mathcal{C}_{2k}) = 0.
\end{align*}
\end{IEEEproof}

\begin{figure*}[htb]
\centering
\includegraphics[scale=0.8]{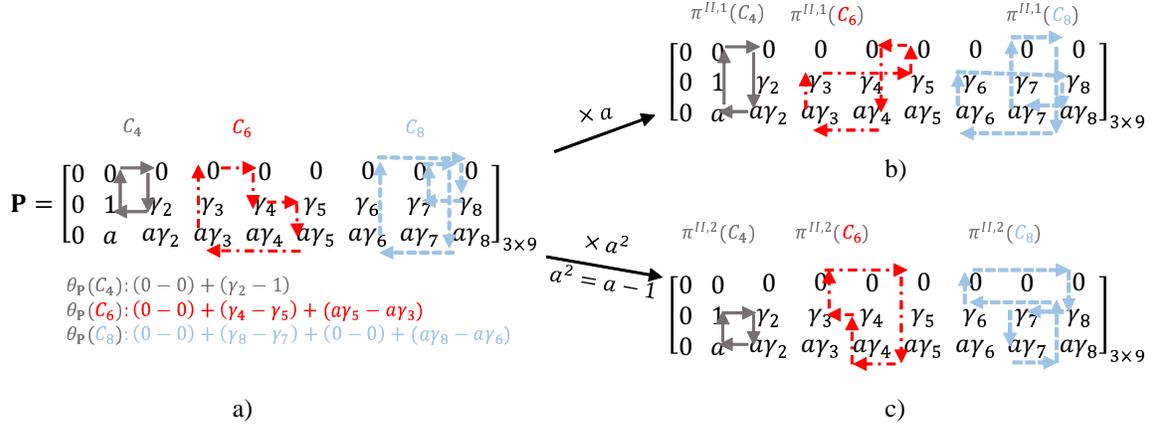}
\caption{Example of cycles of length 4, 6, and 8 and IR-equivalent cycles}
\label{fig:CycleReducdv3}
\end{figure*}

\begin{table*}[htb]
    \centering
    \caption{Computation of $a\lambda(m_{2s}, m_{2s-1})$ for all possible couples.}
    \begin{tabular}{|c|c|c||c|c|} \hline
        $(m_{2s}, m_{2s-1})$ & $\lambda(m_{2s}, m_{2s-1})$  & $a\lambda(m_{2s}, m_{2s-1})$ & $\pi^{II,1}((m_{2s}, m_{2s-1}))$ & $\lambda(\pi^{II,1}(m_{2s}, m_{2s-1})) $ \\ \hline
        $(0,1)$ & $-1$  & $       -a $     & (2,0) &  $ a$ \\
        $(0,2)$ & $-a$  & $-a^2 = 1-a$     & (2,1) &  $a -1$  \\
        $(1,2)$ & $1-a$ & $a-a^2 = 1 $     & (0,1) &  $ -1$ \\
        $(1,0)$ & $1$  & $       a $       & (0,2) &  $-a$ \\
        $(2,0)$ & $a$  & $-a^2 = -1+a$     & (1,2) &  $1 -a$  \\
        $(2,1)$ & $a-1$ & $a-a^2 = -1 $    & (1,0) &  $ 1$ \\  \hline
    \end{tabular}
    \label{typeII}
\end{table*}
\begin{example}
Let $\mathbf{P}_{3\times 9}$ be the exponent matrix of Fig. \ref{fig:CycleReducdv3}. Fig. \ref{fig:CycleReducdv3}.a shows examples of cycles of length 4, 6, and 8. Fig. \ref{fig:CycleReducdv3}.b and \ref{fig:CycleReducdv3}.c show the corresponding IR-equivalent cycles using $\pi^{II,1}$ and $\pi^{II,2}$, respectively. 
\end{example}
Thus, when a type-I or type-II IRS-SMS exponent matrix is used, a new type of equivalence between cycles appears. This equivalence relation will be called IR-equivalence, to distinguish it from the strictly equivalent class, but its effect is identical: it reduces the number of equivalence classes by a factor close to $m-1$ for type-I IRS matrix and close to 3 for type-II IRS matrix. This reduction of the number of equivalence classes translates directly into a reduction of the number of constraints on the exponent matrix. Note that when $m=3$, it is more efficient to search type-II IRS exponent matrices than type-I IRS exponent matrices. In fact, with a type-II IRS exponent matrix, the number of equations is reduced by a factor of 3 whereas is reduced by only a factor $m-1=2$ for a type-I IRS exponent matrix.
\subsection{IRS technique as an {\em a priori} step of greedy search algorithm} \label{subsec: IRS}
Theorems \ref{theorem:cyclic subgroup order m} and \ref{theorem:cyclic subgroup order 3} show that there might exist a proper cyclic subgroup $\langle a\rangle$ of $\mathbb{Z}^{\times}_{N}$ from which we can pick non-zero components of $\vec{P}_1$ as $\vec{P}_1=\left[0,1,a,a^2,\cdots a^{m-2}\right]^{T}$. When $a$ exists, $\vec{P}_1$ is directly defined, and the number of non-equivalent cycles is reduced by at least a factor close to 3. The search algorithm will then take the sub-matrix $\begin{footnotesize}\left[\begin{array}{@{}c | c@{}} \vec{0} & \vec{P}_{1} \end{array}\right]\end{footnotesize}$ as a base and try to find proper values of $\gamma_{j}$s so that $\mathbf{P}_{m\times n}$ meets the girth condition. Nevertheless, for a given $N$ value, there is not always an $a$ value satisfying the condition of theorems \ref{theorem:cyclic subgroup order m} and/or \ref{theorem:cyclic subgroup order 3}. Moreover, even if an $a$ value exists, the resulting two column type-I or type-II exponent matrix $\begin{footnotesize}\left[\begin{array}{@{}c | c@{}} \vec{0} & \vec{P}_{1} \end{array}\right]\end{footnotesize}$ is always of girth 8 but not necessarily of girth 10 or 12. The question is thus which portion of $N$s remains after the sieved process for a given girth $g$ and a given column weight $m$.
\begin{proposition} \label{proposition:two column exponent matrix}
For $N > 3$, all type-II IRS QC-LDPC matrices with an exponent matrix $\mathbf{P}^{a,II}_{3\times 2}$ of size $3\times 2$ are of girth 12.
\end{proposition}
\begin{IEEEproof}
First, the girth is lower than, or equal to, $12$ according to \cite{Tanner1981}. According to (\ref{theorem:nonequivalent potential cycles general case}), there are in $\mathbf{P}^{a,II}_{3 \times 2}$ three cycles of length $4$, no cycle of length $6$, six cycles of length $8$ and no cycle of length $10$. Owing to Theorem \ref{theorem:cyclic subgroup order 3}, the number of cycles to be checked is reduced to one cycle of length 4 and two cycles of length 8. Those cycles can be respectively the cycles $\mathcal{C}_4$, $\mathcal{C}^b_8$ and $\mathcal{C}^c_8$ depicted in parts (b), (e), and (f) of Fig. \ref{fig:CyclePaths}, respectively. Thus,
\begin{align*}
\begin{array}{l}
\theta_\mathbf{P}( \mathcal{C}_4) = 1 \mod N\\
\theta_\mathbf{P}( \mathcal{C}^b_8) = 2 \mod N\\
\theta_\mathbf{P}( \mathcal{C}^c_8) = 2-a \mod N\\
\end{array}
\end{align*} 
If $N>3$, $\theta_\mathbf{P}( \mathcal{C}_4)$  and $\theta_\mathbf{P}( \mathcal{C}^a_8)$ are not equal to zero modulo $N$. The third cycle requires more attention. In fact,  $\theta_\mathbf{P}( \mathcal{C}^b_8) = 0  \mod N$ implies $a = 2$. When $a=2$, $a(1-a) = 1 \mod N$ gives $-3 = 0 \mod N$ and thus $N=3$. Since $N > 3$, $\theta_\mathbf{P}( \mathcal{C}^b_8) \neq 0 \mod N.$
\end{IEEEproof}

For type-I IRS QC-LDPC code, the girth of the QC-LDPC code generated from exponent matrix  $\mathbf{P}^{a,I}_{m \times 2}$, $m>3$ is not always girth $12$. For example, $\mathbf{P}^{a = 73,I}_{4 \times 2}, N=216$ gives a QC-LDPC matrix of girth 8 only, while $\mathbf{P}^{a = 5,I}_{4 \times 2}, N=215$ gives a QC-LDPC matrix of girth 12. Nevertheless, finding a proper $N$ and, accordingly, the existence of a suitable cyclic subgroup that results in $\vec{P}_1$ is not time-consuming. Given a fixed $m$, it will take few milliseconds for MATLAB software to check if $\mathbb{Z}^{\times}_{N}$ is a proper candidate or not. 

\begin{proposition}\label{proposition: different columns type-I}
Let $N\geq 6$, $a,b$ be two different elements of $\mathbb{Z}^{\times}_{N}$ of order $m-1$ with $\langle a\rangle=\langle b\rangle=S$. Owing to row/column permutations, the Tanner graph constructed from type-I exponent matrix $\mathbf{P}^{a,I}_{m\times 2}$ is equivalent to the Tanner graph constructed from type-I exponent matrix $\mathbf{P}^{b,I}_{m\times 2}$.
\end{proposition}

\begin{IEEEproof}
Since $\langle a\rangle=\langle b\rangle=S$, the set of elements of the second column of $\mathbf{P}^{a,I}_{m\times 2}$, i.e. the set $\{0, 1, a, a^2, \ldots, a^{m-2}\}$ is equal to the set of element of the second column $\mathbf{P}^{b,I}_{m\times 2}$, i.e. the set $\{0, 1, b, b^2, \ldots, b^{m-1}\}$. Thus, an appropriate permutation of rows allows the transformation of $\mathbf{P}^{a,I}_{m\times 2}$ into $\mathbf{P}^{b,I}_{m\times 2}$ (the coefficients of the first column are all zero, and thus not affected by any permutation).
\end{IEEEproof}
 
\begin{proposition}\label{proposition: different columns type-II}
Let $N\geq 6$, $a,b$ be two different elements of $\mathbb{Z}^{\times}_{N}$ that satisfy $a(1-a) = 1 \mod N$, $b*(1-b)=1 \mod N$ and $\langle a\rangle=\langle b\rangle=S$. The Tanner graph of the constructed matrix $\mathbf{P}^{a,II}_{m\times 2}$ with the second column $\left[0,1,a\right]$ is equivalent, due to a row permutation, to the Tanner graph of the matrix $\mathbf{P}^{b,II}_{m\times 2}$ with the second column $\left[0,1,b\right]^T$.
\end{proposition}

\begin{IEEEproof} {}The integers $a$ and $b$ verify $a\left(1-a\right)=1=b\left(1-b\right)$, moreover, $O_N(a) = O_N(b) = 6$ (Lemma \ref{lemma:O(6)}). Since a necessary and sufficient condition for non-identity element $z=x^y$ ($\langle x\rangle=S,\;y\in\mathbb{N}$) to be a generator of $S$ is $\gcd\left(y,O(S)\right)=1$, it is easy to see that $a$ and $b=a^5$ are the only generators of $S$. Since, $\gcd(a^5,N)=1$, according to \cite{ATasdighi1},  $a^5\mathbf{P}^{a,II}_{3\times 2}$ and $\mathbf{P}^{a,II}_{3\times 2}$ are equivalent exponent matrices due to row/column permutations. The second column of  $a^5\mathbf{P}^{a,II}_{3\times 2}$ is $\left[0,a^5,1\right]^T = \left[0,b,1\right]^T$. Swapping the second and the third rows of $a^5\mathbf{P}^{a,II}_{3\times 2}$ gives $\mathbf{P}^{b,II}_{3\times 2}$, thus, the two exponent matrices are equivalent.\end{IEEEproof}

 In short, the search algorithm needs to test one generator per each permissible cyclic subgroup $S$ to find the exponent matrix $\mathbf{P}_{m\times n}$ of code with girth $g$ ($g=8, 10,12$).

The final point is ``there might be more than one permissible cyclic subgroup of $\mathbb{Z}^{\times}_{N}$ that meet the conditions to construct type-I or type-II exponent matrices, but not all of them would necessarily result in the matrix $\mathbf{P}_{m\times n}$ with girth $g$ ($g=8, 10,12$)''. For example, $\mathbb{Z}^{\times}_{N=301}$ has two permissible cyclic subgroups $S_1=\langle 80\rangle$ and $S_2=\langle 136\rangle$ of order $6$ where their generators satisfy the property $a(1-a)=1 \mod 301$. The search algorithm in Section \ref{sec:numerical results} can find an exponent matrix $\mathbf{P}^{a=80, II}_{3\times 10}$ with girth $10$ but not with $\mathbf{P}^{a=136, II}_{3\times 10}$.

\begin{figure*}[ht]
    \centering
    \includegraphics{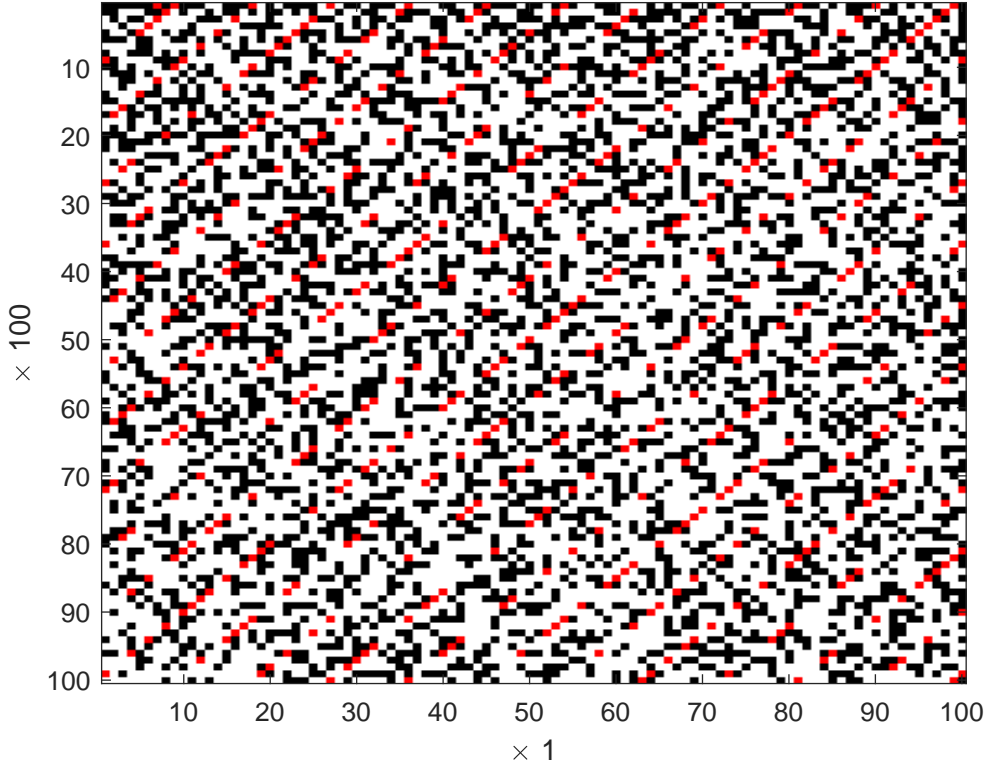}
    \caption{ 2D representation on a $100 \times 100$ grid of the first $10^4$ integers. Integers $N$ having no subgroup of $\mathbb{Z}_N^\times$ of order 4 are represented by black pixels. If no type-I exponent matrix of size (4,2) exists, the integer is represented by a red pixel; it is white otherwise. The first top left pixel corresponds to $N = 1$, and its left neighbor corresponds to $N = 2$. The last pixel of the first line corresponds to $N = 100$. The last bottom right pixel corresponds to $N = 10^{4}$.}
    \label{fig:sievied type-I, m=4}
\end{figure*}

Fig. \ref{fig:sievied type-I, m=4} shows a 2D representation of integers between 1 and $10^4$ sieved by type-I property for $m=4$. The black pixels correspond to values of $N$ with no admissible $a$ value, and the red pixels represent $N$ values with at least one admissible $a$ value but no girth 12 type-I $[\vec{0},\vec{P}_1]$ exponent matrix. Finally, the white pixels indicate $N$ values with at least one girth 12 type-I $[\vec{0},\vec{P}_1]$ exponent matrix. It may be noted that, on average, 60\% of integers between 1 and $10^4$ can give a girth 12 type-I $[\vec{0},\vec{P}_1]$  exponent matrix. We observe that the white pixels are uniformly spread between 1 and $10^4$. For $N$ values corresponding to a white pixel, the number of distinct order 3 subgroups of $\mathbb{Z}_N^\times$ varies from 2 to 13. 
The same uniform dispersion is observed also for type-II QC-LDPC codes and type-I QC-LDPC codes with $m=5$ and $m=6$. The proportion of values able to give a girth 12 two columns IRS exponent matrix are 51.9 \%, 24.2\%, and 13.4\% for type-I $m=5$, type-I $m=6$ and type-II $m=3$, respectively. 

\subsection{Controlled greedy search algorithm}\label{subsec: greedy
search algorithm}
\begin{algorithm}
\caption{Controlled greedy search algorithm for $ m \geq 3 $} \label{Algo1}
\textbf{Input:} \; Parameters $n$, $m$, $N$ of the code, targeted girth
$ g $, vector $ G $ of size $ n $ to control the greedy search effort. \\
\textbf{Output:} \; Eventually, a set of coefficients $ \Gamma_n $ of size $ n $
if success, an empty set otherwise.
\begin{algorithmic}[1]
\StateX \hdashrule[0.5ex]{4cm}{1pt}{1.5mm} Part I: primary step
\hdashrule[0.5ex]{4cm}{1pt}{1.5mm}
\State $ \mathcal{A} \leftarrow \{2, 3, \ldots, N-1 \} $, $\Gamma_n = \varnothing$,
$\Gamma_1 = \{0\}$
\While {$\mathcal{A} \neq\varnothing$ and $ \Gamma_n=\varnothing$}
      \State Extract an element $a$ of $\mathcal{A}$.
      \State $\mathcal{A} \leftarrow \mathcal{A}\setminus\{a\}$
      \If {$O_N(a) = m-1$}
          \State $\vec{P}_1 \leftarrow \left(0,1,a,a^2,\ldots,a^{m-2}\right)^T$
          \State $\mathcal{A} \leftarrow \mathcal{A} \setminus\{a^k\}_{k = 2,3,\ldots ,m-2} $
      \State $\mathcal{S} \leftarrow {\Phi}_g (\Gamma_1, \vec{P}_1, N) $
      \State $ \Gamma_n \leftarrow $ \textbf {search} $ (\Gamma_1, \mathcal{S}, N,
n, \vec{P_1}, G $)
      \EndIf
\EndWhile

\StateX \hdashrule[0.5ex]{4 cm}{1pt}{1.5mm}  Part II: \textbf{search} function
\hdashrule[0.5ex]{6cm}{1pt}{1.5mm}
\State $ \Gamma_n \leftarrow $ \textbf{search} ($ \Gamma, \mathcal{S}, N, n,
\vec{P}_1, G $)
\State $ \Gamma_n \leftarrow \Gamma $
\If {$ | \Gamma_n | = n $}
      Return $ \Gamma_n $
\Else
      \For {$ i = 1 $ to $ | \mathcal{S} | $}
          \State $ s(i) \leftarrow | \mathcal{S} \cap \Phi_g (\Gamma \cup
\mathcal{S}(i), \vec{P}_1, N) |$ (note: $ s $ is a vector).
      \EndFor
      \State $ I \leftarrow $ \textbf{sort\_index} ($s$) (note: $s(I(1)) \geq s(I(2))
\geq \ldots \geq s(I(|\mathcal{S}|)) $).
      \For {$ j = 1 $ to \textbf {min} $ (|\mathcal{S} |, G (| \Gamma |)) $}
         \If {$ | \Gamma_n | = n $}
             Return $ \Gamma_n $
         \Else
             \State $ \Gamma_k \leftarrow \Gamma \cup \{\mathcal{S} (I(j)) \} $
             \State $ \mathcal{S} \leftarrow \mathcal{S} \setminus
\{\mathcal{S} (I(j)) \} $
             \State $\mathcal{S}_k \leftarrow \mathcal{S} \cap \Phi_g (\Gamma_k,
\vec{P}_1, N) $
             \If {$ | \Gamma_k | + | \mathcal{S}_k | \geq n $}
                  \State $ \Gamma_n \leftarrow $ \textbf{search} ($ \Gamma_k,
\mathcal{S}_k, N, n, \vec{P}_1, G $)
             \Else
             \State Return $ \varnothing $
             \EndIf
         \EndIf
      \EndFor
\EndIf
\end{algorithmic}
\end{algorithm}

In this section, we present a modified controlled greedy search algorithm
that uses the SMC technique \cite{MHTAT2018}. 
Let $ \Gamma_k = \{0, 1, \gamma_2, \gamma_3, \ldots, \gamma_{k-1} \} $ be a set of
size $ k $ of elements of $ \mathbb{Z}_{N}$. The property $ \rho_g(
\Gamma_k, \vec{P}_1, N)$ is true if and only if the exponent matrix
$\left[\begin{array} {c | c | c | c | c | c}
\vec{0} & \vec{P}_{1} & \gamma_{2} \otimes \vec{P}_{1} & \gamma_{3} \otimes \vec{P}_{1} & \ldots &
\gamma_{k-1} \otimes \vec{P}_{1} \end{array} \right] $ gives a matrix with
a girth greater than or equal to $ g $ when expanded by a factor of $ N $.
We call $ \Phi_g (\Gamma_{k}, \vec{P}_1, N) $ the ordered set of coefficients of
$ \mathbb{Z}_{N}$ so that a vector $\Gamma_{k + 1}$ of size $k + 1$ constructed
by the concatenation of $ \Gamma_k $ and any coefficient of
$ \Phi_g(\Gamma_k, \vec{P}_1, N)$ also gives an exponent matrix of girth
$ g $. In a more formal way
\begin{equation}
\beta \in \Phi_g (\Gamma_k, \vec{P}_1, N) \iff
\rho_g (\Gamma_k \cup \{\beta\}, \vec{P}_1 , N) \text{ is true}.
\label{eq: rho}
\end{equation}
The search for a solution of degree $ (m, n) $ for a given lifting degree
$N$ is conducted in two steps. The first step involves the enumeration of a single element per class of the $ a $ values satisfying the condition of Theorem \ref{theorem:cyclic subgroup order m}. This step is
described in Algorithm 1 part I for $m> 3$. To do so, the set of values
$\mathcal{A}$ is initialized as $\mathcal{A} = \{2, 3, \ldots, N-1\}$. The values of $\mathcal{A}$ are extracted one by one. The second step starts each time an extracted
value $a$ fulfills the condition of Theorem
\ref{theorem:cyclic subgroup order m}. The function
\textbf{search} is launched to try to find a solution $\Gamma_n$. In
case of success, the algorithm successfully stops. Otherwise, the elements of $ \langle a \rangle $ are suppressed from the search
space $\mathcal{A}$. The process continues until no values remain in
$\mathcal{A}$, and the search is unsuccessful. Note that for
$m=3$, the condition $ O_N(a) = m-1 $ of line 5 should be replaced by the
condition ($a(1-a) = 1 \mod N $), and line $ 7 $ should be replaced by
the instruction $ \mathcal{A} = \mathcal{A} \setminus
\{a^k\}_{k = 1,2,3,4,5}$.

The \textbf{search} function is described in Algorithm 1, part II. It is
a recursive function that tries to increase recursively the size of
$ \Gamma $ until it reaches a size of $ n $. The arguments of the
\textbf{search} function are $ \Gamma $, $ \mathcal{S} $, $ N $, $ n $,
$ \vec{P}_1 $, and a vector $ G $ of size $ n $ that controls the processing
effort. Let us describe the processing during the first call of the
function in line 9. The arguments of this first call are $ \Gamma_1 =
\{0\} $ and $ \mathcal{S} $ (defined in line 8), as the set of values compatible
with $ \Gamma_1 $ (see (\ref{eq: rho})). Lines 14 and 15 set up the greedy
search. For $ i = 1, 2, \ldots, |\mathcal{S} | $, the number $ s (i) $ of
triplets $ \Gamma_3 = \{0, \mathcal{S} (i), \mu \} $, $ \mu \in \mathcal{S} $
satisfying the condition $ \rho(\Gamma_3, \vec{P_1}, N) $ is computed (note that $ s(i) <| \mathcal{S} | $). The $ s(i) $ are thus sorted in decreasing order (line 16), and the first $ G (|\Gamma |) = G (1) $ elements of $ \mathcal{S} $ (line 17) associated with the highest
values of vector $ s $ are tested. For each tested value, a vector $ \Gamma_k $ of size 2 is generated (line 18). The tested value is suppressed from the set $ \mathcal{S} $ (line 19), and then the subset $ \mathcal{S}_k$ of
$\mathcal{S}$ of values compatible with $ \Gamma_k $ is created (line 20).
If the size of $ \mathcal{S}_k $ plus the size of $ \Gamma_k $ is greater than or equal to $ n $, or, if it is still possible to
generate a $ \Gamma $ vector of length $ n $, then the search function is
called again with a $ \Gamma $ set of size $ 2 $. The process is
recursively reiterated until a length $ n $ $ \Gamma $ vector is found or
until no new possibility remains to be explored. The
complexity of the search is controlled by a vector $ G $ of size $ n $. The
$ k^{\text{\small th}} $ value $ G(k) $ of $ G $ indicates that only the most ``promising''
$ G(k) $ branches will be explored inside each depth $k$ recursive call of the search function. Note that when all the values of $G$ are equal to
$ N $ the search algorithm is exhaustive. It can be done in a
limited time (less than a few days) only for low values of $n$. For a large $n$, the first values of $G$ are set to 1 or 2 to reduce the
search space to a reasonable size. Note that $|X|$ represents the
cardinal of the set $X$.
We observe that when taking random coefficients in the exponent matrix $\mathbf{P}$, $\theta_\mathbf{P}(\mathcal{C})$ (see (\ref{eq:Fossorier1})) takes random values between 0 and $N-1$. Thus, the probability that $\theta_\mathbf{P}(\mathcal{C}) \neq 0 \mod N$ is $(1-1/N)$. Assuming that all the cycles are independent (which, of course, is not the case) then, under this groundless hypothesis, the expected number $E_0(m,n,g)$ of girth $g$ $m \times n$ exponent matrices will be given by the size of the space multiplied by the probability that all the cycles satisfy (\ref{eq:Fossorier}), i.e.
\begin{equation}
E_0(m,n,g) = N^{m \cdot n} (1-1/N)^{\#\mathbf{P}^{m \times n}(g)}.
\end{equation}
Using the IRS method, the search space is greatly reduced, but, the number of Fossorier's equation that should be fulfilled is reduced by a factor close to $\max(3, m-1)$. The expectation of finding a solution is thus 
\begin{equation}
E_1(m,n,g) = N^{n-2}  (1-1/N)^{\frac{\#\mathbf{P}^{m \times n}(g)}{\max(3,m-1)}}
\end{equation}
 For $g = 10$, $m=3$, $n=10$, and $N=301$, the ratio $E_0/E_1$ is of order $5\times 10^{-46}$, which shows that finding a girth $g$ code is $10^{46}$ times likely to happen if one uses IRS instead of a normal search, on average. Although the argument is not solid, it helps explain why most of the best IRS matrices found have a smaller size than the already published fully-connected exponent matrices. 

Although the IRS construction method reduces the search space by a factor $N^{m-2}$ compared to the SMC technique alone, the search space is still in $\mathcal{O}(N^{n-2})$.
However, the numerical results presented in the next section demonstrate the efficiency of the proposed IRS construction method.

\section{Numerical Results} \label{sec:numerical results}

\begin{table*}[]
\setlength{\tabcolsep}{.7 pt}
\centering
\caption{Minimum found lifting degree of IRS exponent matrices for girth $g=8,10$ and 12. $N_{min}^{g}$ is the smallest found lifting degree of the exponent matrix with girth $g$. The lifting degree of the shortest existing codes is given with an exponent that indicates the corresponding reference. Exponents $^{a, b, c, d, e, f, g, h}$ refer to \cite{Bocharova1}, \cite{ATasdighi1}, \cite{MBATMB2018}, \cite{AEA2018}, \cite{HHDBH2018}, \cite{Xu2019Tanner},  \cite{sullivan1}, \cite{ATasdighi2} respectively.}

 \begin{tabular}{|c|c|c|c|c|c|c|c|c|c|c|c|c|c|c|} \hline
           &
           $\mathbf{P}_{3\times 4}^{a,II}$   &      $\mathbf{P}_{3\times 5}^{a,II}$   &      $\mathbf{P}_{3\times 6}^{a,II}$   & $\mathbf{P}_{3\times 7}^{a,II}$   & $\mathbf{P}_{3\times 8}^{a,II}$   & $\mathbf{P}_{3\times 9}^{a,II}$   & $\mathbf{P}_{3\times 10}^{a,II}$  & $\mathbf{P}_{3\times 11}^{a,II}$  & $\mathbf{P}_{3\times 12}^{a,II}$  & $\mathbf{P}_{3\times 13}^{a,II}$  & $\mathbf{P}_{3\times 14}^{a,II}$  & $\mathbf{P}_{3\times 15}^{a,II}$  & $\mathbf{P}_{3\times 16}^{a,II}$ &$\mathbf{P}_{3\times 17}^{a,II}$    \\  \hline
            \multirow{2}{*}{$N_{min}^{g=10}$} &
          37 & 61 & 91 & 133 & 181 & 241 & 301 & 373 & 463 & 571   & 727  &  877 & 1039 & 1231 \\
          &$37^{a,b}$& $61^{a,b}$& $91^{b}$& $139^{c,d}$ & $181^{c}$   & $241^{c}$   &  $313^{c}$ & $397^{c}$  & $523^{c}$ & - & -   & -  &  - & - \\ 
          \hline
          \multirow{2}{*}{$N_{min}^{g=12}$} &
          73 & 151 & 271 & 427 & 619 & 921 & 1303 & 2011 & 2883 & 3769 & 4953 & 6321 & - & -\\
          & $73^{a,b}$   & $151^{c}$   &  $271^{c}$ & $457^{c}$  & $691^{c}$ & $991^{c}$ & $1447^{c}$   & $2161^{c}$  &  $4730^{a}$ & $5851^{e}$ &- &- &- &- \\ \hline
          &
          $\mathbf{P}_{3\times 18}^{a,II}$  & $\mathbf{P}_{3\times 19}^{a,II}$  &  $\mathbf{P}_{3\times 20}^{a,II}$  & $\mathbf{P}_{3\times 21}^{a,II}$  & $\mathbf{P}_{3\times 22}^{a,II}$  & $\mathbf{P}_{3\times 23}^{a,II}$  &  $\mathbf{P}_{3\times 24}^{a,II}$  & $\mathbf{P}_{3\times 25}^{a,II}$  & $\mathbf{P}_{3\times 26}^{a,II}$  & $\mathbf{P}_{3\times 27}^{a,II}$  & $\mathbf{P}_{3\times 28}^{a,II}$  & $\mathbf{P}_{3\times 29}^{a,II}$  & $\mathbf{P}_{3,30}^{a,II}$ & $\mathbf{P}_{3,31}^{a,II}$   \\ \hline
        \multirow{2}{*}{$N_{min}^{g=10}$} &
            1453 & 1723 & 2089 & 2197 & 2689   & 3049  &  3331 &  3577 & - & - & - & - & - & -\\ 
            &-&-&-&-&-&$5659^{f}$ & - & $4801^{f}$ &- &- &- &- &- &- \\
\hline
\hline
           &
           $\mathbf{P}_{4\times 4}^{a,I}$   &      $\mathbf{P}_{4\times 5}^{a,I}$   &      $\mathbf{P}_{4\times 6}^{a,I}$   & $\mathbf{P}_{4\times 7}^{a,I}$   & $\mathbf{P}_{4\times 8}^{a,I}$   & $\mathbf{P}_{4\times 9}^{a,I}$   & $\mathbf{P}_{4\times 10}^{a,I}$  & $\mathbf{P}_{4\times 11}^{a,I}$  & $\mathbf{P}_{4\times 12}^{a,I}$  & $\mathbf{P}_{4\times 13}^{a,I}$  & $\mathbf{P}_{4\times 14}^{a,I}$  & $\mathbf{P}_{4\times 15}^{a,I}$  & $\mathbf{P}_{4\times 16}^{a,I}$ &$\mathbf{P}_{4\times 17}^{a,I}$    \\  \hline
            \multirow{2}{*}{$N_{min}^{g=10}$} &
            73 & 133 & 199 & 247 & 403 & 541 & 703 & 883 & 1123 & 1429 & 1933 & 2389 & 2881 & 3397 \\
            & - &
            $139^{c}$ & $241^{c}$ & $307^{c}$ & $409^{c}$ & $577^{c}$ & $787^{c}$ & $1039^{c}$ & $1381^{c}$ & - & - & - & - & - \\ \hline
            \multirow{2}{*}{$N_{min}^{g=12}$} &
             254 & 571 & 1087 & 2203 & 4489 & 8966 & - & - & - & - & - &   - & - & - 
            \\
            & - &  
            $607^{c}$ & $1201^{c}$ & $2371^{c}$ & $6607^{g}$ & $12071^{g}$ & - & - & - & -  & - & - & - & -  \\ \hline \hline 
           &
           $\mathbf{P}_{5\times 4}^{a,I}$   &      $\mathbf{P}_{5\times 5}^{a,I}$   &      $\mathbf{P}_{5\times 6}^{a,I}$   & $\mathbf{P}_{5\times 7}^{a,I}$   & $\mathbf{P}_{5\times 8}^{a,I}$   & $\mathbf{P}_{5\times 9}^{a,I}$   & $\mathbf{P}_{5\times 10}^{a,I}$  & $\mathbf{P}_{5\times 11}^{a,I}$  & $\mathbf{P}_{5\times 12}^{a,I}$  & $\mathbf{P}_{5\times 13}^{a,I}$  & $\mathbf{P}_{5\times 14}^{a,I}$  & $\mathbf{P}_{5\times 15}^{a,I}$  & $\mathbf{P}_{5\times 16}^{a,I}$ &$\mathbf{P}_{5\times 17}^{a,I}$    \\  \hline
 \multirow{2}{*}{$N_{min}^{g=10}$} &
            175 & 205 & 511 & 763 & 1067 & 1417 & 1903 & 2431 & 3445 & 4849 & 5933 & - & - & - \\
            & - & - & -  & $1471^{f}$ & - & $1621^{f}$ & - & $2861^{f}$ & - & $5981^{f}$ & - & - & - & \\ 
            \hline \hline
           &
           $\mathbf{P}_{6\times 4}^{a,I}$   &      $\mathbf{P}_{6\times 5}^{a,I}$   &      $\mathbf{P}_{6\times 6}^{a,I}$   & $\mathbf{P}_{6\times 7}^{a,I}$   & $\mathbf{P}_{6\times 8}^{a,I}$   & $\mathbf{P}_{6\times 9}^{a,I}$   & $\mathbf{P}_{6\times 10}^{a,I}$  & $\mathbf{P}_{6\times 11}^{a,I}$  & $\mathbf{P}_{6\times 12}^{a,I}$  & $\mathbf{P}_{6\times 13}^{a,I}$  & $\mathbf{P}_{6\times 14}^{a,I}$  & $\mathbf{P}_{6\times 15}^{a,I}$  & $\mathbf{P}_{6\times 16}^{a,I}$ & $\mathbf{P}_{6\times 17}^{a,I}$   \\  \hline
            \multirow{2}{*}{$N_{min}^{g=8}$} &
            41 & 61 & 101 & 101 & 121 & 151 & 181 & 181 & 181 & 241   & 281  &  331 & 341 & 401 \\
          &-& -& -& $70^{h}$   & $95^{h}$   &  $125^{h}$ & $150^{h}$  & $182^{h}$ & $218^{h}$ & $254^{h}$   & $296^{h}$  &  $337^{h}$ & $380^{h}$ & $429^{h}$\\ \hline
          $N_{min}^{g=10}$ &
            251 & 421 & 571 & 971 & 1331 & 1891 & 2621 & 3421 & 4261 & 5611   & 7171  &  - & - & - \\ \hline
          &
          $\mathbf{P}_{6\times 18}^{a,I}$  & $\mathbf{P}_{6\times 19}^{a,I}$  &  $\mathbf{P}_{6\times 20}^{a,I}$  & $\mathbf{P}_{6\times 21}^{a,I}$  & $\mathbf{P}_{6\times 22}^{a,I}$  & $\mathbf{P}_{6\times 23}^{a,I}$  &  $\mathbf{P}_{6\times 24}^{a,I}$  & $\mathbf{P}_{6\times 25}^{a,I}$  & $\mathbf{P}_{6\times 26}^{a,I}$  & $\mathbf{P}_{6\times 27}^{a,I}$  & $\mathbf{P}_{6\times 28}^{a,I}$  & $\mathbf{P}_{6\times 29}^{a,I}$  & $\mathbf{P}_{6,30}^{a,I}$ &  $\mathbf{P}_{6,31}^{a,I}$\\ \hline
            \multirow{2}{*}{$N_{min}^{g=8}$} &
            451 & 521 & 571 & 601 & 661   & 751  &  781 &  881 & 941 & 941 & 1051 & 1111 & 1111 & -\\ 
           &$478^{g}$ & $530^{g}$  & $584^{g}$ & - & $2113^{f}$   & $967^{f}$  &  -  & $1951^{f}$  & $2029^{f}$  &  - & $3529^{f}$ & $4003^{f}$ & - &- \\ \hline
\end{tabular}
\label{table:dv3456g81012}
\end{table*}

Table \ref{table:dv3456g81012} shows, for several sets of parameters $(m,n,g)$, the minimum lifting degree $N_{min}$ found to allow constructing a type-I and type-II IRS QC-LDPC matrix (the description of the found IRS matrices is shown in Appendix I). In Table \ref{table:dv3456g81012}, the best state-of-the-art minimum $N$ value is also provided with the associated reference. 

A first conclusion of this study is that the state-of-the-art lower bound on the lifting degree required to obtain a girth-10 expanded matrix \cite{FAmirzade1} is a little over-estimated when $m>3$ and $n>3$. In fact, we found a counter example for $(m,n) = (4,7)$. The lower bound is supposed to be $L(m,n) = 2\binom{m}{2}\binom{n}{2}+1$, which gives, $L(4,7) = 253$ whereas we found a girth-10 $4 \times 7$ expanded matrix with a lifting degree of $N = 247$ (see Appendix I). A corrected lower bound derived from the Double Difference matrix defined in \cite{FAmirzade1} is derived in Appendix II. The corrected lower bound $L^c(m,n)$ is given as
\begin{equation}
    L^c(m,n) = 2\binom{m}{2}\binom{n}{2} - 2\binom{m-2}{2}\binom{n-2}{2} +1.
    \label{eq:L_c}
\end{equation}
Note that $L^c(4,7) = 237$, which is compatible with the found value $N = 247$.

A second conclusion is that we were always able to find an equal or better (sometimes, significantly better) solution than that of the state of the art for girth 10 and girth 12 matrices. For girth 8, no matrices of variable node degree $m < 6$ better than the state-of-the-art have been found. Hence, the result are not reported. For $m=6$, except for $n \in \{7,8,9,10\}$, better solutions have been found.
Finally, future search simulation may help us find smaller lifting degree. Updated values will be reported online in \cite{LDPCLABSTICC}.
\section{Conclusion}
A new construction method of fully-connected QC-LDPC codes of girth $g = 8$, $10$, and $12$ has been presented. This method is called Integer Ring Sieve. For a lifting factor of size $N$, the IRS construction requires that the second column of the exponent matrix takes its values in an integer ring of $\mathbb{Z}_N^\times$ with a specific property. The constraint on the second column has several beneficial effects: it reduces the search space, it reduces the number of constraints to be checked, and ultimately, for most of the exponent matrix size $(m,n)$  with a girth objective of 8, 10, or 12, it gives a smaller lifting factor than the ones already reported in the literature. The table of best-found IRS QC-LDPC codes is also given in the paper. Finally, the IRS method allows us to find a counter-example showing that the lower bound proposed in \cite{ FAmirzade1} is a little over-estimated. A corrected lower bound has been given.
\section*{Acknowledgment}
This project has been partly funded by the Brittany Region in the frame of the call ``Campagne d'Activit\'e Postdoctorale 2017''. The authors would like to thank the anonymous reviewers and the associated editor, M. Andrew Thangaraj, for their insightful comments that helped to improve the paper.
\section*{Appendix I}
In this appendix, an abstract list of the parameters of our constructed codes is provided. The parameters of each code are embedded in braces as $\{N, \mathbf{P}^{a,T}_{m \times n},  [0,1,\gamma_2,\gamma_3,\cdots,\gamma_{n-1}] \}$, with $T$ indicating the type of IR exponent matrix, i.e. $\mathbf{P}^{a,I}$ or $\mathbf{P}^{a,II}$.

\begin{small}
 \underline{Case $m=3$, girth 10: }  
$\{N = 37, \: \mathbf{P}^{27, II}_{3 \times 4}$, [0, 1, 3, 24]\}; 
$\{N = 61, \: \mathbf{P}^{14, II}_{3 \times 5}$, [0, 1, 3, 21, 55]\}; 
$\{N = 91, \: \mathbf{P}^{17, II}_{3 \times 6}$, [0, 1, 3, 7, 25, 38]\}; 
$\{N = 133, \: \mathbf{P}^{12, II}_{3 \times 7}$, [0, 1, 3, 32, 38, 42, 116]\}; 
$\{N = 181, \: \mathbf{P}^{133, II}_{3 \times 8}$, [0, 1, 3, 69, 120, 129, 141, 156]\}; 
$\{N = 241, \: \mathbf{P}^{16, II}_{3 \times 9}$, [0, 1, 3, 13, 88, 114, 182, 217, 223]\}; 
$\{N = 301, \: \mathbf{P}^{80, II}_{3 \times 10}$, [0, 1, 3, 7, 33, 73, 117, 140, 208, 226]\}; 
$\{N = 373, \: \mathbf{P}^{285, II}_{3 \times 11}$, [0, 1, 3, 35, 50, 73, 95, 170, 180, 221, 235]\}; 
$\{N = 463, \: \mathbf{P}^{442, II}_{3 \times 12}$, [0, 1, 3, 9, 29, 116, 148, 219, 260, 329, 388, 418]\}; 
$\{N = 571, \: \mathbf{P}^{462, II}_{3 \times 13}$, [0, 1, 3, 9, 91, 120, 140, 217, 375, 398, 511, 516, 561]\}; 
$\{N = 727, \: \mathbf{P}^{446, II}_{3 \times 14}$, [0, 1, 3, 7, 12, 35, 105, 192, 213, 352, 442, 472, 653, 714]\}; 
$\{N = 877, \: \mathbf{P}^{595, II}_{3 \times 15}$, [0, 1, 3, 7, 12, 22, 47, 114, 247, 390, 423, 431, 639, 692, 755]\}; 
$\{N = 1039, \: \mathbf{P}^{899, II}_{3 \times 16}$, [0, 1, 3, 7, 12, 20, 36, 183, 396, 462, 674, 716, 798, 823, 967, 982]\}; 
$\{N = 1231, \: \mathbf{P}^{1105, II}_{3 \times 17}$, [0, 1, 3, 7, 12, 20, 34, 106, 132, 374, 402, 450, 519, 737, 1010, 1061, 1071]\}; 
$\{N = 1453, \: \mathbf{P}^{760, II}_{3 \times 18}$, [0, 1, 3, 7, 12, 20, 30, 46, 132, 184, 239, 320, 418, 867, 951, 1015, 1100, 1382]\}; 
$\{N = 1723, \: \mathbf{P}^{1682, II}_{3 \times 19}$, [0, 1, 3, 7, 12, 20, 30, 46, 67, 99, 248, 605, 693, 793, 831, 975, 1105, 1271, 1381]\}; 
$\{N = 2089, \: \mathbf{P}^{1263, II}_{3 \times 20}$, [0, 1, 3, 7, 12, 20, 30, 45, 61, 85, 107, 249, 510, 602, 970, 1022, 1297, 1481, 1635, 1987]\}; 
$\{N = 2197, \: \mathbf{P}^{1161, II}_{3 \times 21}$, [0, 1, 122, 125, 251, 303, 413, 493, 811, 846, 867, 877, 1262, 1416, 1438, 1533, 1739, 1794, 2083, 2109, 2191]\}; 
$\{N = 2689, \: \mathbf{P}^{2298, II}_{3 \times 22}$, [0, 1, 17, 39, 66, 196, 432, 466, 522, 524, 1109, 1217, 1257, 1343, 1596, 1788, 1998, 2255, 2346, 2504, 2524, 2618]\}; 
$\{N = 3049, \: \mathbf{P}^{2517, II}_{3 \times 23}$, [0, 1, 89, 267, 414, 586, 612, 639, 710, 726, 1002, 1373, 1424, 1504, 1573, 1821, 1971, 2077, 2145, 2338, 2445, 2646, 2886]\}; 
$\{N = 3331, \: \mathbf{P}^{1868, II}_{3 \times 24}$, [0, 1, 31, 242, 399, 404, 407, 557, 716, 916, 1209, 1623, 1843, 1878, 1941, 1998, 2013, 2229, 2318, 2436, 2618, 2676, 3139, 3231]\}; 
$\{N = 3577, \: \mathbf{P}^{1452, II}_{3 \times 25}$, [0, 1, 7, 91, 164, 571, 580, 674, 677, 1033, 1070, 1346, 1657, 2443, 2651, 2700, 2772, 2895, 2916, 2931, 3049, 3144, 3204, 3467, 3523]\}; 

 \underline{Case $m=3$, girth 12: }  
$\{N = 73, \: \mathbf{P}^{9, II}_{3 \times 4}$, [0, 1, 3, 13]\}; 
$\{N = 151, \: \mathbf{P}^{119, II}_{3 \times 5}$, [0, 1, 3, 108, 139]\}; 
$\{N = 271, \: \mathbf{P}^{29, II}_{3 \times 6}$, [0, 1, 3, 7, 67, 144]\}; 
$\{N = 427, \: \mathbf{P}^{136, II}_{3 \times 7}$, [0, 1, 3, 18, 209, 300, 388]\}; 
$\{N = 619, \: \mathbf{P}^{367, II}_{3 \times 8}$, [0, 1, 3, 216, 312, 318, 462, 529]\}; 
$\{N = 921, \: \mathbf{P}^{632, II}_{3 \times 9}$, [0, 1, 3, 117, 226, 232, 384, 441, 595]\}; 
$\{N = 1303, \: \mathbf{P}^{1208, II}_{3 \times 10}$, [0, 1, 5, 14, 89, 349, 383, 562, 1130, 1152]\}; 
$\{N = 2011, \: \mathbf{P}^{1806, II}_{3 \times 11}$, [0, 1, 3, 10, 30, 122, 454, 654, 937, 1095, 1699]\}; 
$\{N = 2883, \: \mathbf{P}^{2444, II}_{3 \times 12}$, [0, 1, 11, 442, 522, 902, 965, 1145, 1857, 2091, 2632, 2775]\}; 
$\{N = 3769, \: \mathbf{P}^{3306, II}_{3 \times 13}$, [0, 1, 19, 154, 1257, 1539, 1636, 2519, 2564, 2855, 3099, 3111, 3250]\}; 
$\{N = 4953, \: \mathbf{P}^{1544, II}_{3 \times 14}$, [0, 1, 108, 457, 486, 1252, 1331, 1546, 2558, 3839, 4262, 4308, 4746, 4911]\}; 
$\{N = 6321, \: \mathbf{P}^{2273, II}_{3 \times 15}$, [0, 1, 827, 1613, 1637, 2135, 3891, 4051, 4082, 4342, 4380, 4694, 5171, 5328, 5905]\}; 

 \underline{Case $m=4$, girth 10: }  
$\{N = 73, \: \mathbf{P}^{8, I}_{4 \times 4}$, [0, 1, 34, 47]\}; 
$\{N = 133, \: \mathbf{P}^{11, I}_{4 \times 5}$, [0, 1, 5, 21, 54]\}; 
$\{N = 199, \: \mathbf{P}^{92, I}_{4 \times 6}$, [0, 1, 3, 104, 147, 161]\}; 
$\{N = 247, \: \mathbf{P}^{68, I}_{4 \times 7}$, [0, 1, 83, 206, 209, 215, 220]\}; 
$\{N = 403, \: \mathbf{P}^{87, I}_{4 \times 8}$, [0, 1, 3, 7, 111, 159, 233, 303]\}; 
$\{N = 541, \: \mathbf{P}^{129, I}_{4 \times 9}$, [0, 1, 3, 99, 264, 314, 353, 401, 423]\}; 
$\{N = 703, \: \mathbf{P}^{26, I}_{4 \times 10}$, [0, 1, 9, 123, 353, 443, 498, 501, 609, 663]\}; 
$\{N = 883, \: \mathbf{P}^{545, I}_{4 \times 11}$, [0, 1, 16, 210, 471, 589, 652, 776, 780, 824, 877]\}; 
$\{N = 1123, \: \mathbf{P}^{1089, I}_{4 \times 12}$, [0, 1, 56, 144, 330, 695, 733, 829, 848, 987, 994, 1112]\}; 
$\{N = 1429, \: \mathbf{P}^{764, I}_{4 \times 13}$, [0, 1, 73, 217, 263, 269, 461, 805, 918, 1020, 1253, 1297, 1396]\}; 
$\{N = 1933, \: \mathbf{P}^{1341, I}_{4 \times 14}$, [0, 1, 13, 297, 299, 370, 445, 576, 871, 985, 1277, 1376, 1735, 1886]\}; 
$\{N = 2389, \: \mathbf{P}^{1699, I}_{4 \times 15}$, [0, 1, 28, 63, 203, 450, 506, 1413, 1455, 1471, 1478, 1544, 1640, 2008, 2040]\}; 
$\{N = 2881, \: \mathbf{P}^{2315, I}_{4 \times 16}$, [0, 1, 464, 786, 831, 931, 1032, 1090, 1111, 1326, 1929, 2136, 2164, 2177, 2239, 2275]\}; 
$\{N = 3397, \: \mathbf{P}^{2788, I}_{4 \times 17}$, [0, 1, 3, 23, 154, 606, 647, 861, 1432, 1496, 1636, 1922, 2455, 2699, 2786, 3215, 3300]\}; 

 \underline{Case $m=4$, girth 12: }  
$\{N = 254, \: \mathbf{P}^{107, I}_{4 \times 4}$, [0, 1, 25, 46]\}; 
$\{N = 571, \: \mathbf{P}^{461, I}_{4 \times 5}$, [0, 1, 17, 184, 482]\}; 
$\{N = 1087, \: \mathbf{P}^{829, I}_{4 \times 6}$, [0, 1, 4, 142, 1018, 1055]\}; 
$\{N = 2203, \: \mathbf{P}^{1917, I}_{4 \times 7}$, [0, 1, 4, 130, 443, 1082, 1397]\}; 
$\{N = 4489, \: \mathbf{P}^{3789, I}_{4 \times 8}$, [0, 1, 942, 1062, 1312, 1547, 2202, 3692]\}; 
$\{N = 8966, \: \mathbf{P}^{3977, I}_{4 \times 9}$, [0, 1, 11, 17, 1158, 2049, 3754, 4987, 6942]\}; 

 \underline{Case $m=5$, girth 10: }  
$\{N = 175, \: \mathbf{P}^{118, I}_{5 \times 4}$, [0, 1, 6, 165]\}; 
$\{N = 205, \: \mathbf{P}^{9, I}_{5 \times 5}$, [0, 1, 4, 52, 193]\}; 
$\{N = 511, \: \mathbf{P}^{265, I}_{5 \times 6}$, [0, 1, 3, 114, 244, 354]\}; 
$\{N = 763, \: \mathbf{P}^{251, I}_{5 \times 7}$, [0, 1, 3, 72, 81, 117, 720]\}; 
$\{N = 1067, \: \mathbf{P}^{604, I}_{5 \times 8}$, [0, 1, 3, 8, 32, 46, 812, 1050]\}; 
$\{N = 1417, \: \mathbf{P}^{142, I}_{5 \times 9}$, [0, 1, 3, 225, 386, 912, 972, 1294, 1337]\}; 
$\{N = 1903, \: \mathbf{P}^{439, I}_{5 \times 10}$, [0, 1, 395, 464, 500, 988, 1139, 1350, 1686, 1877]\}; 
$\{N = 2431, \: \mathbf{P}^{395, I}_{5 \times 11}$, [0, 1, 493, 510, 698, 832, 1091, 1370, 1433, 1867, 1979]\}; 
$\{N = 3445, \: \mathbf{P}^{818, I}_{5 \times 12}$, [0, 1, 10, 200, 281, 461, 583, 1364, 1408, 1641, 2178, 2492]\}; 
$\{N = 4849, \: \mathbf{P}^{3626, I}_{5 \times 13}$, [0, 1, 136, 218, 392, 732, 1048, 1244, 2143, 2642, 3476, 3522, 4024]\}; 
$\{N = 5933, \: \mathbf{P}^{5099, I}_{5 \times 14}$, [0, 1, 484, 1320, 1457, 1738, 2287, 2691, 3651, 3696, 3891, 4065, 4715, 5375]\}; 

 \underline{Case $m=6$, girth 8: }  
$\{N = 41, \: \mathbf{P}^{10, I}_{6 \times 4}$, [0, 1, 2, 3]\}; 
$\{N = 61, \: \mathbf{P}^{9, I}_{6 \times 5}$, [0, 1, 2, 12, 13]\}; 
$\{N = 101, \: \mathbf{P}^{36, I}_{6 \times 6}$, [0, 1, 2, 3, 4, 5]\}; 
$\{N = 101, \: \mathbf{P}^{36, I}_{6 \times 7}$, [0, 1, 2, 3, 4, 26, 79]\}; 
$\{N = 121, \: \mathbf{P}^{3, I}_{6 \times 8}$, [0, 1, 2, 7, 12, 24, 64, 116]\}; 
$\{N = 151, \: \mathbf{P}^{8, I}_{6 \times 9}$, [0, 1, 2, 3, 4, 5, 6, 49, 108]\}; 
$\{N = 181, \: \mathbf{P}^{42, I}_{6 \times 10}$, [0, 1, 2, 3, 10, 11, 21, 25, 101, 173]\}; 
$\{N = 181, \: \mathbf{P}^{42, I}_{6 \times 11}$, [0, 1, 2, 3, 10, 35, 41, 51, 77, 132, 173]\}; 
$\{N = 181, \: \mathbf{P}^{42, I}_{6 \times 12}$, [0, 1, 3, 10, 35, 38, 52, 86, 89, 105, 147, 156]\}; 
$\{N = 241, \: \mathbf{P}^{87, I}_{6 \times 13}$, [0, 1, 2, 4, 5, 16, 31, 104, 106, 125, 175, 208, 212]\}; 
$\{N = 281, \: \mathbf{P}^{86, I}_{6 \times 14}$, [0, 1, 13, 14, 18, 116, 117, 122, 146, 149, 173, 178, 179, 277]\}; 
$\{N = 331, \: \mathbf{P}^{64, I}_{6 \times 15}$, [0, 1, 2, 20, 34, 61, 88, 100, 108, 116, 123, 158, 186, 201, 216]\}; 
$\{N = 341, \: \mathbf{P}^{47, I}_{6 \times 16}$, [0, 1, 12, 44, 83, 84, 98, 101, 115, 116, 155, 187, 198, 199, 222, 318]\}; 
$\{N = 401, \: \mathbf{P}^{39, I}_{6 \times 17}$, [0, 1, 15, 27, 31, 36, 51, 54, 58, 65, 76, 90, 105, 111, 112, 113, 145]\}; 
$\{N = 451, \: \mathbf{P}^{16, I}_{6 \times 18}$, [0, 1, 9, 109, 117, 133, 152, 157, 158, 159, 192, 207, 255, 263, 292, 314, 330, 401]\}; 
$\{N = 521, \: \mathbf{P}^{25, I}_{6 \times 19}$, [0, 1, 2, 99, 100, 101, 196, 201, 234, 297, 298, 299, 320, 409, 421, 434, 462, 470, 475]\}; 
$\{N = 571, \: \mathbf{P}^{106, I}_{6 \times 20}$, [0, 1, 37, 103, 205, 225, 252, 262, 273, 274, 275, 296, 299, 317, 349, 368, 424, 509, 547, 570]\}; 
$\{N = 601, \: \mathbf{P}^{32, I}_{6 \times 21}$, [0, 1, 27, 50, 69, 136, 137, 138, 144, 238, 279, 325, 412, 463, 488, 495, 502, 531, 589, 596, 600]\}; 
$\{N = 661, \: \mathbf{P}^{197, I}_{6 \times 22}$, [0, 1, 10, 24, 56, 57, 58, 59, 82, 113, 117, 152, 176, 251, 331, 360, 412, 474, 565, 590, 603, 605]\}; 
$\{N = 751, \: \mathbf{P}^{80, I}_{6 \times 23}$, [0, 1, 11, 45, 51, 124, 210, 211, 212, 318, 342, 355, 407, 420, 421, 542, 550, 551, 579, 612, 621, 629, 750]\}; 
$\{N = 781, \: \mathbf{P}^{5, I}_{6 \times 24}$, [0, 1, 2, 29, 150, 151, 152, 154, 210, 216, 223, 235, 255, 268, 300, 358, 364, 369, 379, 483, 502, 517, 598, 630]\}; 
$\{N = 881, \: \mathbf{P}^{268, I}_{6 \times 25}$, [0, 1, 2, 106, 204, 226, 319, 326, 327, 328, 363, 366, 391, 416, 430, 537, 548, 554, 555, 556, 645, 656, 657, 678, 735]\}; 
$\{N = 941, \: \mathbf{P}^{349, I}_{6 \times 26}$, [0, 1, 40, 59, 80, 202, 227, 228, 229, 255, 347, 419, 453, 457, 466, 484, 488, 491, 531, 684, 712, 713, 714, 715, 847, 940]\}; 
$\{N = 941, \: \mathbf{P}^{349, I}_{6 \times 27}$, [0, 1, 3, 53, 96, 202, 227, 228, 229, 243, 260, 321, 340, 381, 419, 453, 457, 460, 484, 630, 644, 707, 712, 713, 714, 715, 940]\}; 
$\{N = 1051, \: \mathbf{P}^{307, I}_{6 \times 28}$, [0, 1, 3, 72, 73, 74, 75, 114, 116, 134, 136, 218, 222, 248, 251, 371, 447, 493, 614, 655, 696, 737, 770, 887, 977, 979, 981, 985]\}; 
$\{N = 1111, \: \mathbf{P}^{339, I}_{6 \times 29}$, [0, 1, 25, 90, 100, 156, 182, 191, 203, 211, 302, 303, 312, 351, 381, 406, 529, 619, 628, 685, 731, 755, 762, 763, 799, 810, 833, 851, 1088]\}; 
$\{N = 1111, \: \mathbf{P}^{339, I}_{6 \times 30}$, [0, 1, 25, 90, 100, 156, 182, 191, 203, 211, 302, 303, 312, 351, 381, 406, 529, 619, 628, 685, 731, 755, 762, 763, 799, 810, 833, 851, 1088, 1107]\}; 

 \underline{Case $m=6$, girth 10: }  
$\{N = 142, \: \mathbf{P}^{5, I}_{6 \times 3}$, [0, 1, 48]\}; 
$\{N = 251, \: \mathbf{P}^{20, I}_{6 \times 4}$, [0, 1, 4, 79]\}; 
$\{N = 421, \: \mathbf{P}^{252, I}_{6 \times 5}$, [0, 1, 4, 94, 387]\}; 
$\{N = 571, \: \mathbf{P}^{106, I}_{6 \times 6}$, [0, 1, 45, 154, 272, 382]\}; 
$\{N = 971, \: \mathbf{P}^{803, I}_{6 \times 7}$, [0, 1, 6, 585, 608, 725, 872]\}; 
$\{N = 1331, \: \mathbf{P}^{1170, I}_{6 \times 8}$, [0, 1, 8, 45, 145, 674, 970, 1292]\}; 
$\{N = 1891, \: \mathbf{P}^{95, I}_{6 \times 9}$, [0, 1, 124, 394, 534, 1025, 1076, 1224, 1236]\}; 
$\{N = 2621, \: \mathbf{P}^{1295, I}_{6 \times 10}$, [0, 1, 29, 375, 392, 446, 846, 1643, 2014, 2048]\}; 
$\{N = 3421, \: \mathbf{P}^{36, I}_{6 \times 11}$, [0, 1, 203, 1000, 1247, 1478, 1645, 1660, 2245, 2466, 2526]\}; 
$\{N = 4261, \: \mathbf{P}^{398, I}_{6 \times 12}$, [0, 1, 723, 1697, 2055, 2405, 2533, 2574, 2737, 3334, 3398, 3879]\}; 
$\{N = 5611, \: \mathbf{P}^{783, I}_{6 \times 13}$, [0, 1, 202, 421, 2209, 3070, 3556, 3616, 3726, 3869, 4068, 4261, 4537]\}; 
$\{N = 7171, \: \mathbf{P}^{238, I}_{6 \times 14}$, [0, 1, 248, 703, 735, 936, 1304, 2618, 3613, 4332, 4353, 4848, 5360, 6771]\}; 
\end{small}
\section*{Appendix II}
In this appendix, we give first a toy counter-example showing that condition (ii) of Theorem 1 of \cite{FAmirzade1} is not a necessary condition to prevent girth-8 cycles, and thus, that the derived lower bound is over-estimated. Next, we proposed an updated accurate lower bound.

Let us consider the following exponent matrix $\mathbf{P}_{4\times 4}$ with an expansion factor $N$, where $\infty$ denotes the zero matrix. 
\begin{equation}
    \mathbf{P}_{4\times 4} = \begin{bmatrix}
0 & 0 & \infty & \infty \\
0 & a & \infty & \infty \\
\infty & \infty & 0 & 0 \\
\infty & \infty & 0 & b 
\end{bmatrix}
\end{equation}
The corresponding $\binom{4}{2} \times \binom{4}{2} = 6 \times 6$ double difference $\mathbf{DD}$ matrix defined in \cite{FAmirzade1} is given as
\begin{equation}
    \mathbf{DD} = \begin{bmatrix}
(a,N-a) & (\infty,\infty) &\ldots& (\infty,\infty) & (\infty,\infty) \\
(\infty,\infty)  & (\infty,\infty) &\ldots& (\infty,\infty) & (\infty,\infty) \\
(\infty,\infty)  & (\infty,\infty) &\ldots& (\infty,\infty) & (\infty,\infty) \\
(\infty,\infty)  & (\infty,\infty) &\ldots& (\infty,\infty) & (\infty,\infty) \\
(\infty,\infty)  & (\infty,\infty) &\ldots& (\infty,\infty) & (\infty,\infty) \\
(\infty,\infty)  & (\infty,\infty) &\ldots& (\infty,\infty) & (b,N-b) \\
\end{bmatrix}
\end{equation}
In \cite{FAmirzade1}, the lower bound is derived from the claim that all the finite values of the $L(m,n) = 2\binom{m}{2}\binom{n}{2}$ elements of $\mathbf{DD}$ should be distinct. In the toy counter-example, the double difference matrix $\mathbf{DD}$ of $\mathbf{P}_{4 \times 4}$ contains 4 finite values, which gives a lower bound equals to $4 + 1 = 5$. Nevertheless, setting $a = 1, b = 1$ and $N=3$ gives a matrix of girth 12, which is in contradiction with the fact that the lower bound is equal to $N=5$. The trick is that two elements of the matrix $\mathbf{DD}$ corresponding to two length-4 cycles that share neither lines nor columns can be equal without generating a length-8 cycle in the expanded matrix.

From this observation, it is possible to update the lower bound for a fully connected exponent matrix $\mathbf{P}_{m \times n}$. Let $\mathbf{DD}$ be the double difference matrix associated to $\mathbf{P}_{m \times n}$ \cite{FAmirzade1}. The couple $\mathbf{DD}(1,1)$ is equal to $\mathbf{DD}(1,1) = (\theta_\mathbf{P}(\mathcal{C}_4^0), -\theta_\mathbf{P}(\mathcal{C}_4^0)\mod N)$ with $\mathcal{C}_4^0$ defined as the cycle that goes through the first two lines and the first two rows of $\mathbf{P}_{m\times n}$, i.e., 
$\theta_\mathbf{P}(\mathcal{C}_4^0) = p_{1,1} - p_{1, 2} + p_{2,2} - p_{2,1} \mod N$ (see Fig. \ref{fig:CyclePaths}.b). Any length-4 cycles generated from two columns with an index ranging from 3 to $n$ and two lines with an index ranging from 3 to $m$ are constructed as completely disjoint from the cycle $\mathcal{C}_4^0$. Thus, there are exactly $E(m,n) = \binom{n-2}{2} \binom{m-2}{2}$ other length-4 cycles in $\mathbf{P}_{m \times n}$ that share neither row nor column with $\mathcal{C}_4^0$. Those $E(m,n)$ length-4 cycles are associated with $E(m,n)$ 
positions in the $\mathbf{DD}$ matrix, and consequently with $2E(m,n)$ values that can be equal either to $\theta_\mathbf{P}(\mathcal{C}_4^0)$ or to $-\theta_\mathbf{P}(\mathcal{C}_4^0) \mod N$ without generating a length-8 cycle. The lower bound for a fully connected QC-LDPC code is thus corrected as $L^c(m,n) = L(m,n) - 2E(m,n) + 1$ i.e.
\begin{equation}
L^c(m,n) = 2\binom{n}{2} \binom{m}{2} -  2\binom{n-2}{2} \binom{m-2}{2} + 1.
\end{equation}
\bibliographystyle{IEEEtran}
\bibliography{mybib}
\begin{IEEEbiographynophoto}{Alireza Tasdighi}
received the M.A.Sc. Degree in applied mathematics from Sharif University of Technology, Tehran, Iran in 2011. In 2011-2012, he was instructor at the Department of Basic Sciences, Persian Golf University, Bushehr, Iran. From 2012 to 2016, he pursued his Ph.D. in information/coding theory at Amirkabir University of Technology, Tehran, Iran. As a part of his Ph.D. program, Alireza joined a channel coding research group at Department of System and Computer Engineering, Carleton University, Ottawa, Canada. From 2017 to 2018, Alireza has been a faculty member of the IASBS university (Zanjan, Iran). In 2019, he was a posdoct researcher at LAB-STICC, Universit\'e Bretagne Sud (Lorient, France). Since, 2020, he is a postdoc researcher at the  MEE Department at IMT Atlantique (Brest, France). Alireza's interests include graph theory, number theory, linear algebra, protograph based QC-LDPC codes, Non-Binary QC-LDPC codes, remote sensing and machine learning.
\end{IEEEbiographynophoto}
\begin{IEEEbiographynophoto}{Emmanuel Boutillon}
received the Engineering Diploma in 1990 and its Ph.D. degree in 1995, both from the Telecom Paris Tech, Paris. From 1995 to 2000, he was an assistant professor in Telecom Paris Tech. In 1998, he spent a sabbatical year at the University of Toronto, Ontario, Canada. In 2000, he moved to the Universit\'e Bretagne Sud (Lorient, France) as a professor. He headed the LESTER lab from 2005 up to the end of 2007. He was then head of CACS department (lab-STICC) until 2016. In 2011, he had a sabbatical year at INICTEL-UNI, Lima (Peru). His research interests are on the interactions between algorithm and architecture in the field of wireless communications and high speed signal processing. In particular, he works on binary and non-binary decoders.
\end{IEEEbiographynophoto}
\end{document}